\documentclass[english,usenatbib]{mn2e}
\usepackage{mn2e-breakabs}
\usepackage{url}
\usepackage[a4paper,margin=0.8in]{geometry}
\usepackage{aas_macros}
\usepackage[english]{babel}
\usepackage[utf8x]{inputenc}
\usepackage{algorithmic}
\usepackage{algorithm}
\usepackage{units}
\usepackage{amsmath}
\usepackage{graphicx}

\def\pvm#1{[PM: {\it #1}] }
\def\pvm2#1{}

\newcommand{\be}{\begin{equation}}
\newcommand{\ee}{\end{equation}}
\newcommand{\ba}{\begin{eqnarray}}
\newcommand{\ea}{\end{eqnarray}}
\newcommand{\mperh}{\,h^{-1}\,{\rm Mpc}}
\newcommand{\hperm}{\,h\,{\rm Mpc}^{-1}}

\usepackage[breaklinks, colorlinks, citecolor=blue, linkcolor=black, urlcolor=black]{hyperref}

\begin{document}
\title[mock generation for galaxy surveys]
{
Accurate halo-galaxy mocks from automatic bias estimation and particle mesh gravity solvers 
}
\author[Vakili et al.]{
  \parbox{\textwidth}{
Mohammadjavad~Vakili$^1$\thanks{E-mail: mjvakili@nyu.edu},
Francisco-Shu~Kitaura$^2$,$^3$\thanks{E-mail: fkitaura@iac.es},
Yu~Feng$^4$,\\
Gustavo Yepes$^5$,
Cheng~Zhao$^6$,
Chia-Hsun~Chuang$^7$,
ChangHoon~Hahn$^1$
}
  \vspace*{10pt} \\
$^1$ Center for Cosmology and Particle Physics, Department of Physics, New York University, New York, NY, 10003, USA\\
$^2$ Instituto de Astrofísica de Canarias, 38205 San Cristóbal de La Laguna, Santa Cruz de Tenerife, Spain\\
$^{3}$ Departamento de Astrof{\'i}sica, Universidad de La Laguna (ULL), E-38206 La Laguna, Tenerife, Spain\\
$^4$ Berkeley Center for Cosmological Physics, Department of Physics, University of California Berkeley, Berkeley CA, 94720, USA\\
$^5$ Departamento de F{\'i}sica Te{\'o}rica,  Universidad Aut{\'o}noma de Madrid, Cantoblanco, 28049, Madrid, Spain\\ 
$^6$ Tsinghua Center for Astrophysics, Department of Physics, Tsinghua University, Haidian District, Beijing 100084, P. R. China\\ 
$^7$ Leibniz-Institut f\"ur Astrophysik Potsdam (AIP), An der Sternwarte 16, D-14482 Potsdam, Germany\\ 
}

\date{\today} 

\maketitle
\begin{abstract}
Reliable extraction of cosmological information from clustering measurements of galaxy surveys requires estimation of the error covariance matrices of observables. The accuracy of covariance matrices is limited by our ability to generate sufficiently large number of independent mock catalogs that can describe the physics of galaxy clustering across a wide range of scales. 
Furthermore, galaxy mock catalogs are required to study systematics in galaxy surveys and to test analysis tools.
In this investigation, we present a fast and accurate approach for generation of mock catalogs for the upcoming galaxy surveys. 
Our method relies on low-resolution approximate gravity solvers to simulate the large scale dark matter field, which we then populate with halos according to a flexible nonlinear and stochastic bias model. 
In particular, we extend the \textsc{patchy} code with an efficient particle mesh algorithm to simulate the dark matter field (the \textsc{FastPM} code), and with a robust MCMC method relying on the \textsc{emcee} code for constraining the parameters of the bias model. 
Using the halos in the BigMultiDark high-resolution $N$-body simulation as a reference catalog, we demonstrate that our technique can model the bivariate probability distribution function (counts-in-cells), power spectrum, and bispectrum of halos in the reference catalog. Specifically, we show that the new ingredients permit us to reach percentage accuracy in the power spectrum up to $k\sim 0.4\; \hperm$ (within 5\% up to $k\sim 0.6\; \hperm$) with accurate bispectra improving previous results based on Lagrangian perturbation theory.
\end{abstract}


\section{Introduction}

The current and the next generation of galaxy surveys such as \textsc{eBOSS}\footnote{\url{http://www.sdss.org/surveys/eboss/}} (Extended Baryon Oscillation Spectroscopic Survey, \citealt{eBOSS}), \textsc{DESI}\footnote{\url{http://desi.lbl.gov/}} (Dark Energy Spectroscopic Instrument, \citealt{desi}), \textsc{EUCLID}\footnote{\url{http://www.euclid-ec.org/}} (\citealt{euclid}), \textsc{LSST}\footnote{\url{http://www.lsst.org/}} (\citealt{lsst}), and \textsc{WFIRST}\footnote{\url{https://www.nasa.gov/wfirst}} (\citealt{wfirst}) are expected to achieve unprecedented constraints on the cosmological parameters, growth of structure, expansion history of the universe, and modified theories of gravity. Accurate cosmological inferences with these surveys require accurate computation of the likelihood function of the observed data given a cosmological model. This goal can be achieved provided that the uncertainties, in the form of error covariance matrices in the likelihood functions, are reliably estimated. Therefore, covariance matrices are essential ingredients in extraction of cosmological information from the data.

The most commonly used technique in estimation of the covariance matrix for galaxy clustering observables requires generation of a large number of simulated galaxy mock catalogs. These mock catalogs need to reproduce the cosmic volume probed by the galaxy surveys. They also need to describe the clustering observables with high accuracy in a wide range of scales. It has been demonstrated that both the precision and the accuracy of constraints on the cosmological parameters, regardless of the details of a given galaxy survey, depend on the number of realizations of the survey (\citealt{dodelson2013,taylor2014}). The requirement on the number of independent realizations of the survey becomes more stringent as the number of data points in a given analysis grows (\citealt{taylor2013}).
The most pressing challenges ahead of simulating a large number of catalogs are: simulation of large volumes for sampling the Baryonic Acoustic feature in the galaxy clustering, accurate description of the clustering signal at small scales, accurate clustering not only at the level of two-point statistics but also at the level of higher order statistics, and resolving low mass halos that host fainter galaxy samples.  

High-resolution $N$-body simulations are ideal venues for reproducing the dark matter clustering accurately. But production of a large number of density field realizations with $N$-body simulations is not computationally feasible. In order to alleviate the computational cost of $N$-body simulations, several methods based on approximate gravity solvers have been introduced. Methods based on higher order Lagrangian perturbation theory (\citealt{buchert1993,bouchet1995,catelan1995,monaco2002,scocci2002,alpt}), Zeldovich approximation (\citealt{eazymock}), and approximate $N$-body simulations (\citealt{cola2013,qpm,howlet2015,cola,fastpm,ice_cola,koda}) have been demonstrated to be promising for fast generation of dark matter density field. Sampling the structures such as galaxies and halos from the dark matter density field requires an additional step. Identification of virialized regions of matter overdensity is either done through a biasing scheme (\citealt{kitaura2014,qpm}) or is done through application of friends-of-friends algorithm (\citealt{pthalo,koda,fastpm}). Methods that employ a biasing scheme need to be calibrated such that they are statistically consistent with accurate $N$-body simulations or observations. 

The \textsc{patchy} method (\citealt{kitaura2014,kitaura2015}) produces mock catalogs by first generating dark matter field with Lagrangian Perturbation Theory modified with spherical collapse model on small scales ($r \leq 2 \; \mperh$) and then sampling galaxies (halos) from the density field using nonlinear stochastic biasing introduced in \citet{kitaura2014}. This method has been shown to reproduce the two-point clustering with $\sim$2\% accuracy down to $k \sim 0.3 \; \hperm$ and the counts-in-cells of the massive halos in an accurate $N$-body simulation. \citet{kitaura2015} demonstrate that the mock catalogs generated using this technique are capable of accurately describing the halo bispectrum in the reference $N$-body simulations. Furthermore, \citet{kitaura2016} used this method for massive production of mock catalogs for the cosmological analysis of the completed SDSS III Baryon Oscillation Spectroscopic Survey DR12 galaxy sample. 



Alternatively, error covariance matrices can be computed with analytical models (\citealt{feldman1994,smith2008,crocce2011,sun2013,grieb2016,klaus2016}).
These methods are promising, though still need further investigation especially including systematic effects, such as the survey geometry. They will potentially permit us to use a smaller number of mock catalogs to obtain accurate covariance matrices.

In recent years, development of the shrinkage methods (\citealt{ledoit2004,pope2008,ledoit2012,joachimi2016,simpson2016}) have been proved promising for alleviating the requirement on the number of mocks. In principle, one could use a combination of the shrinkage methods and a smaller number of mock catalogs to reach the same level of accuracy needed for large scale structure inferences.    

Moreover, production of mocks will be a useful tool for investigation of possible sources of systematic errors as well as verification of covariance matrices derived from analytical methods.

In this investigation, we introduce an MCMC method for calibration of the bias model of the \textsc{patchy} code. This method constrains the bias parameters by the halo power spectrum and the halo counts-in-cells (halo PDF) of a reference halo catalog constructed from an accurate $N$-body simulation. 

Furthermore, we replace the dark matter gravity solver of the code with the fast particle-mesh approximate $N$-body solver implemented in the \textsc{FastPM} code (\citealt{fastpm}). The advantage of the \textsc{FastPM} algorithm over other methods based on particle-mesh is its low memory requirements as well as accurate large scale growth. In addition, the dark matter density field produced by the \textsc{FastPM} code yields better nonlinear clustering than that of the perturbation theory. 

As a proof of concept, we make use of the halos in the BigMultiDark Planck high-resolution $N$-body simulation (\citealt{multidark}). This catalog has been extensively used for validation, comparison and production of galaxy mock catalogs (\citealt{chuang2015,zhao2015,kitaura2016,sergio2016}). 
In addition, we will make a statistical comparison between our \textsc{patchy} mocks and the reference catalog. We present the number density, halo PDF (halo counts-in-cells), and halo two-point statistics. 
We also present our results in terms of the three-point statistics since it is rising as a major complementary approach in various large-scale structure analyses \citep{slepian2015,gill2015a,gill2015b,guo2016,slepian2016a,slepian2016b,gill2017}.

The remainder of this paper is structured as follows: In section \S \ref{sec:method}, we present our method for generating and calibrating mock catalogs. This includes description of the structure formation model, nonlinear stochastic bias model of the \textsc{patchy} code, and our MCMC method for constraining the bias parameters. In section \S \ref{sec:code_integration}, we describe how different pieces of code are integrated into the \textsc{patchy} method. We illustrate the performances using a reference halo catalog constructed from an accurate $N$-body simulation in section \S \ref{sec:test}, and we discuss the main results and present our conclusions in section \S \ref{sec:discussion}.

\section{Methodology}
\label{sec:method}

Our method consists of producing the large scale dark matter field on a mesh and then populating it with halos (or galaxies) with a given bias model. The parameters of that bias model are constrained with a reference catalog in an automatic statistical way. 
Our approach is agnostic about the method used for identification of halos in the reference catalog. The \textsc{patchy} code permits us to sample galaxies directly from the density field. For instance \citet{kitaura2016} samples mock galaxy catalogs based on an accurate reference mock galaxy catalog \citep{sergio2016}. 
Let us first describe in \S \ref{sec:sf} the new implementation of the structure formation in \textsc{patchy}, followed in \S \ref{sec:bias} by the bias model, and finally in  \S \ref{sec:mcmc} our novel MCMC sampling procedure to obtain the bias parameters. 

\begin{figure*}
 \begin{tabular}{ccc}
\includegraphics[width=0.6\columnwidth]{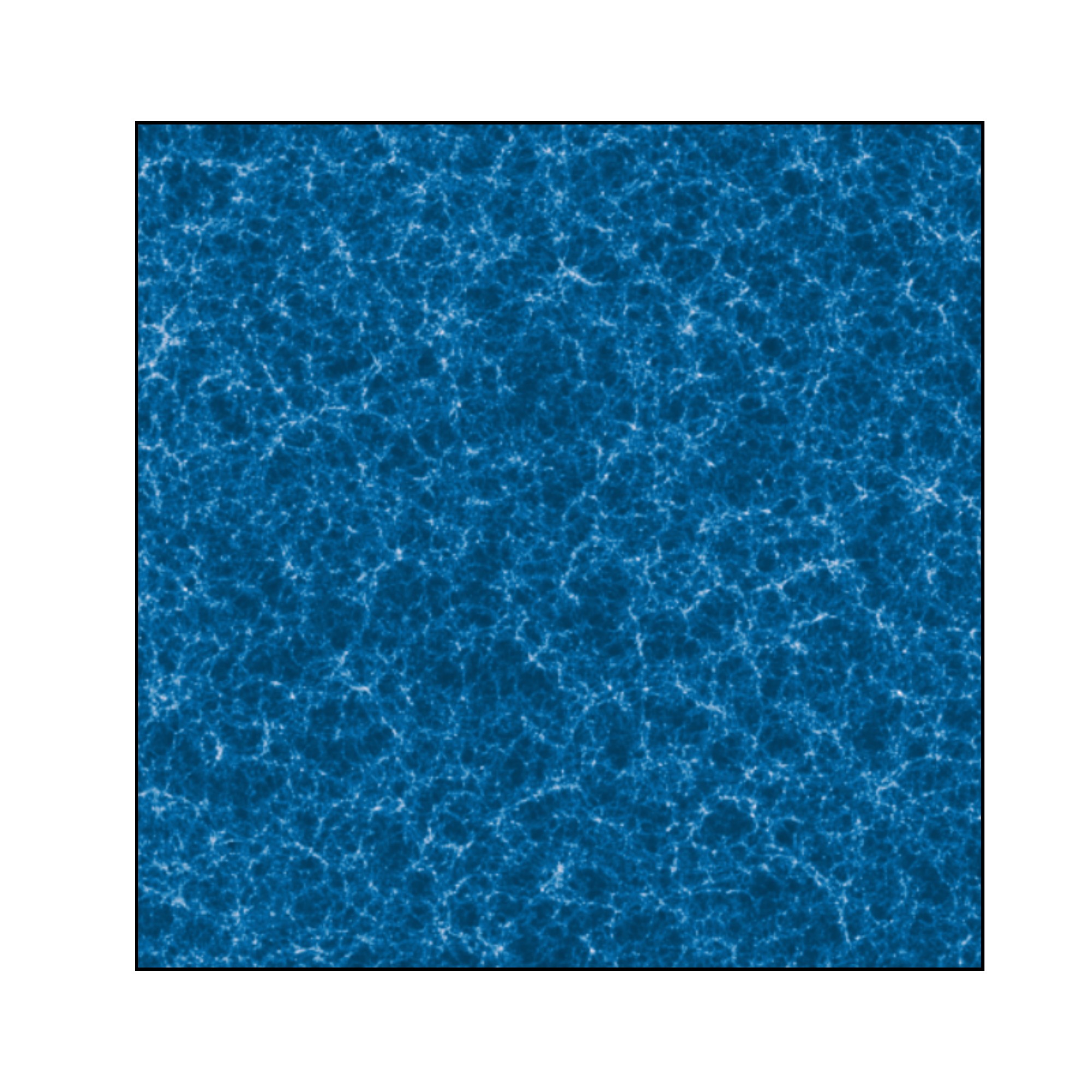}
\includegraphics[width=0.6\columnwidth]{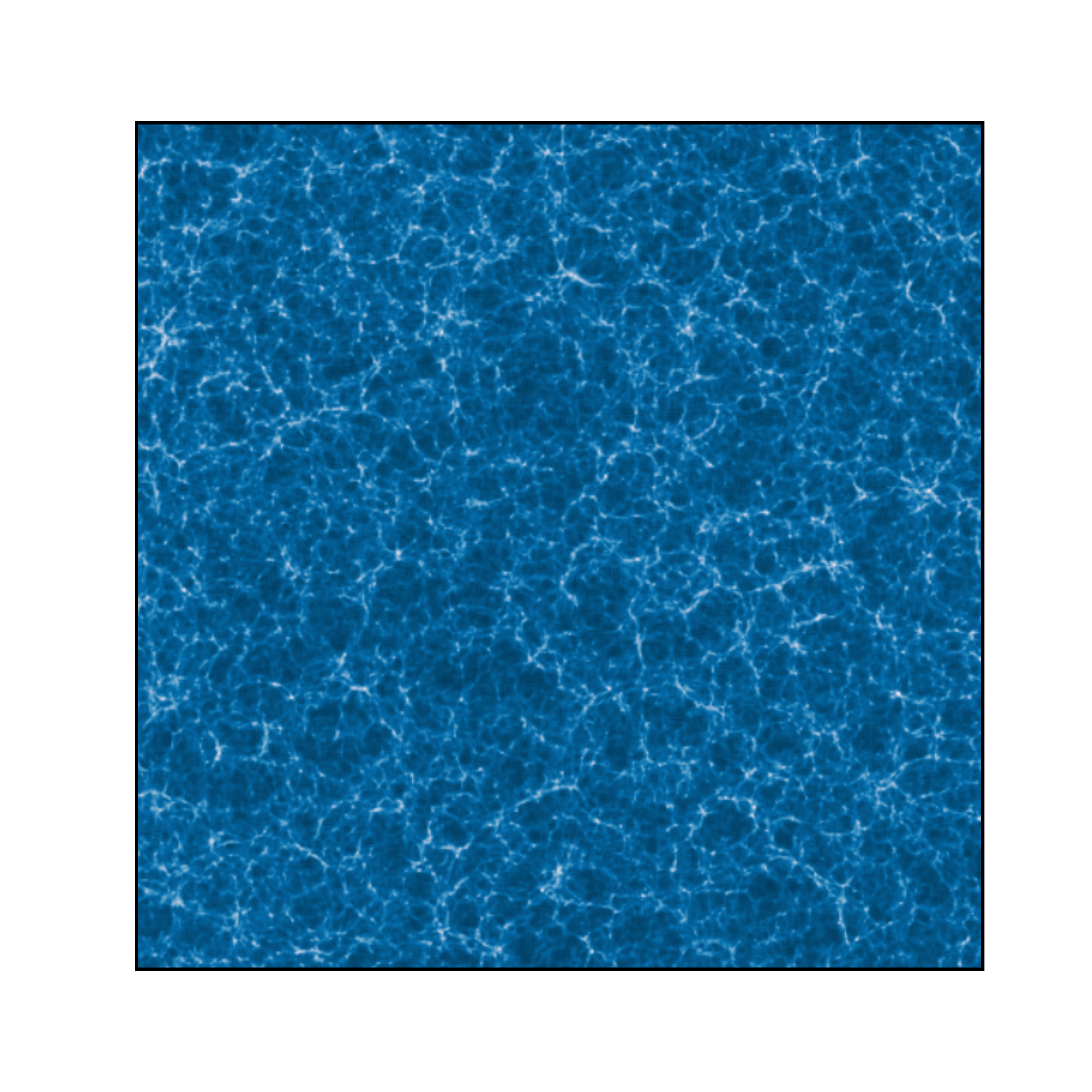}
\includegraphics[width=0.6\columnwidth]{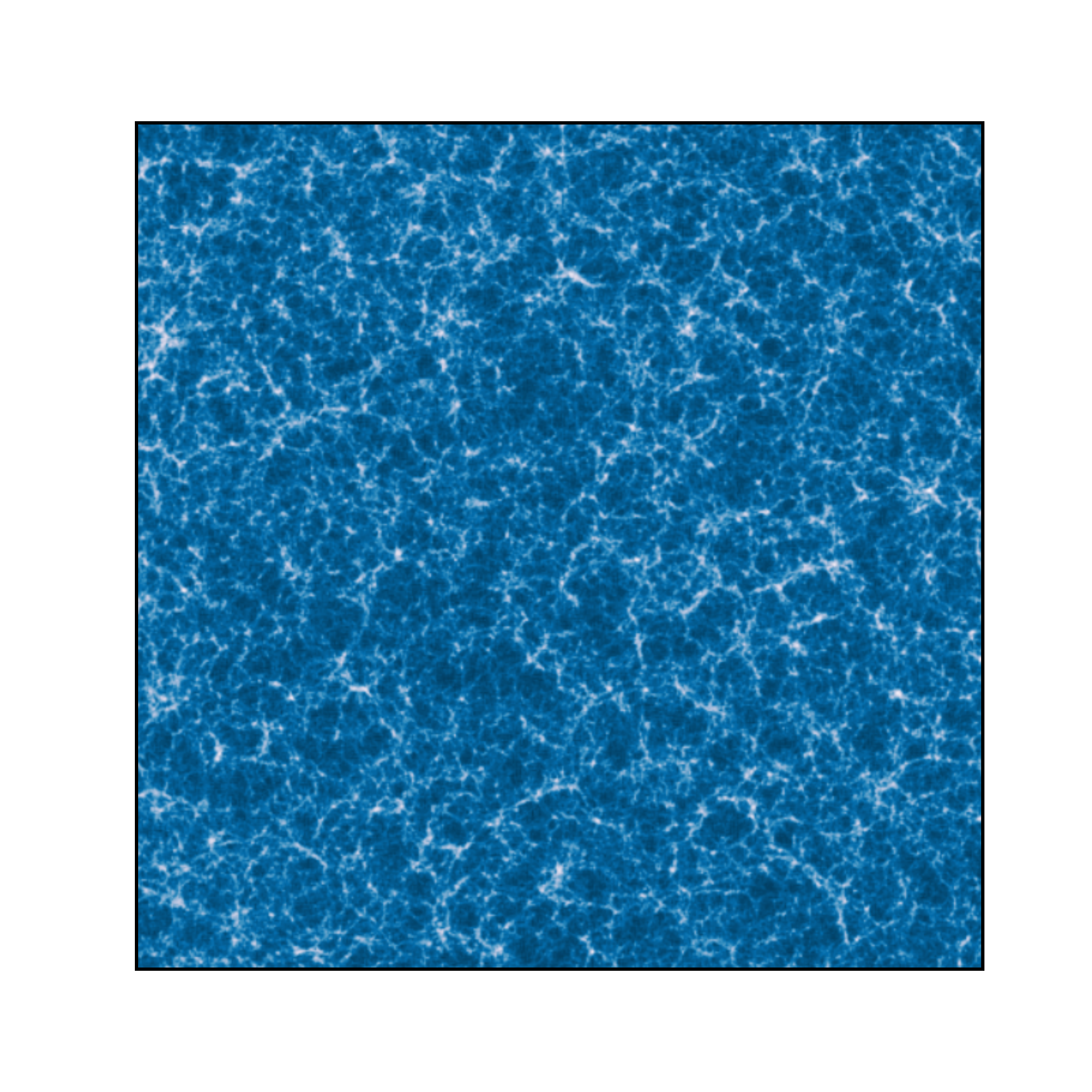}\\
\includegraphics[width=0.6\columnwidth]{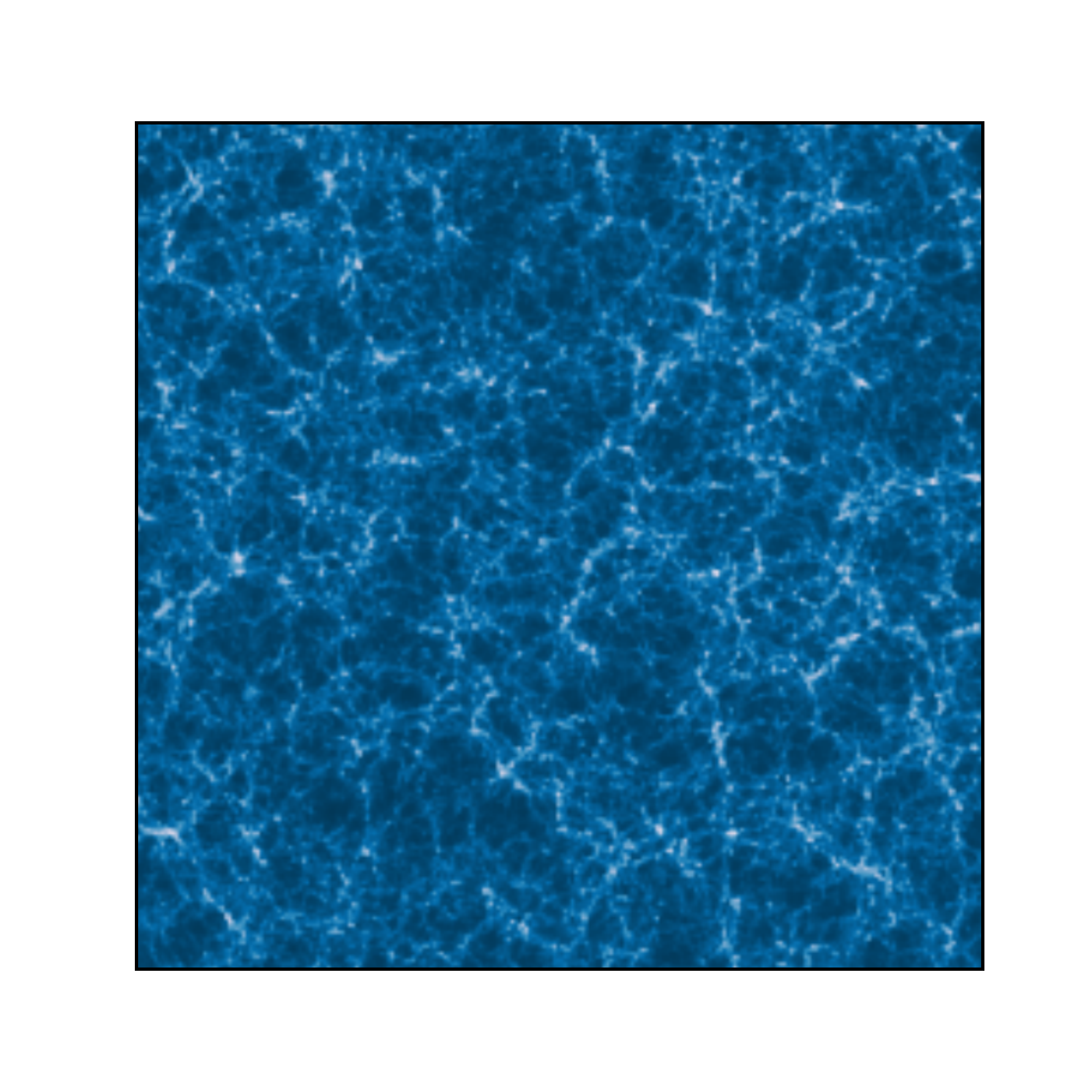}
\includegraphics[width=0.6\columnwidth]{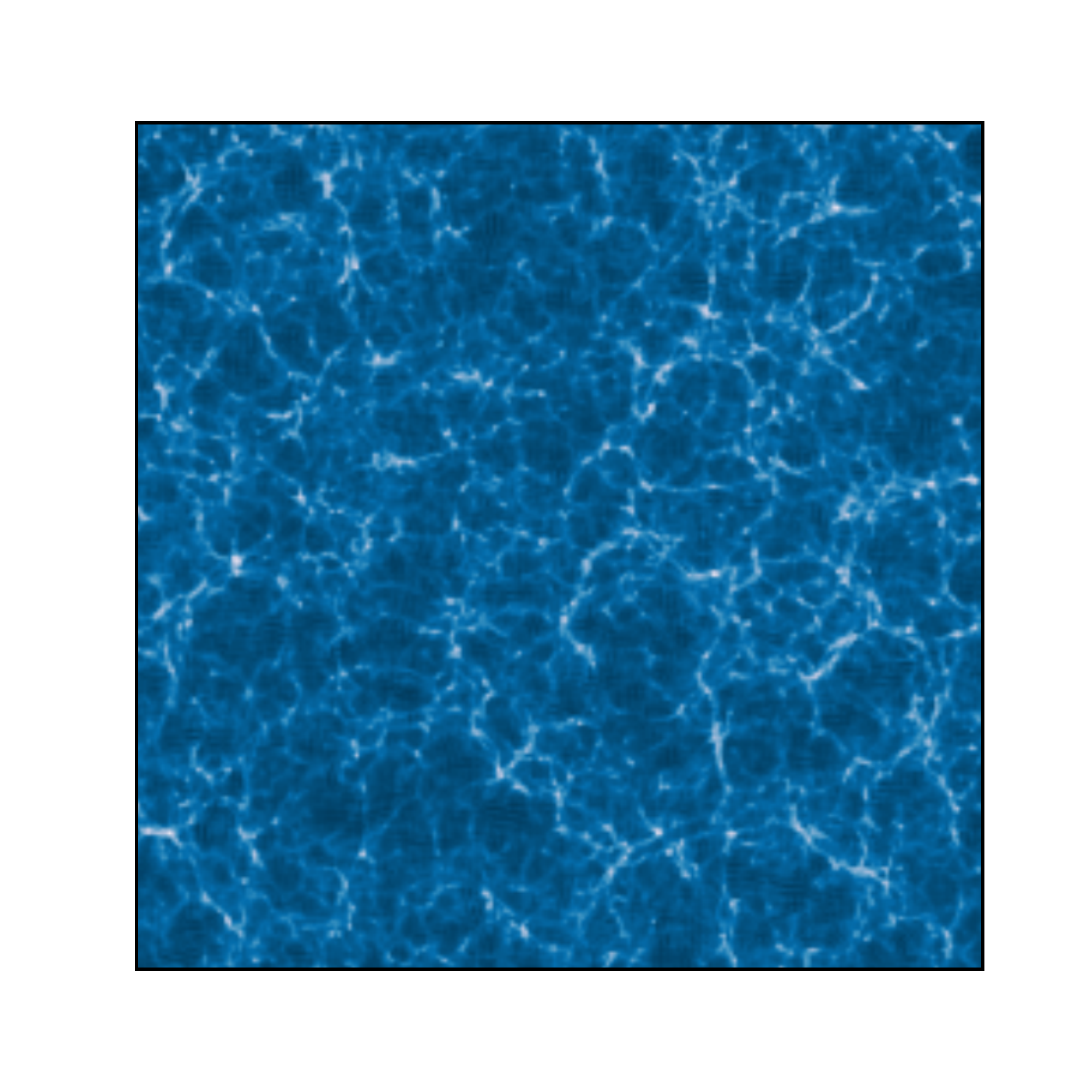}
\includegraphics[width=0.6\columnwidth]{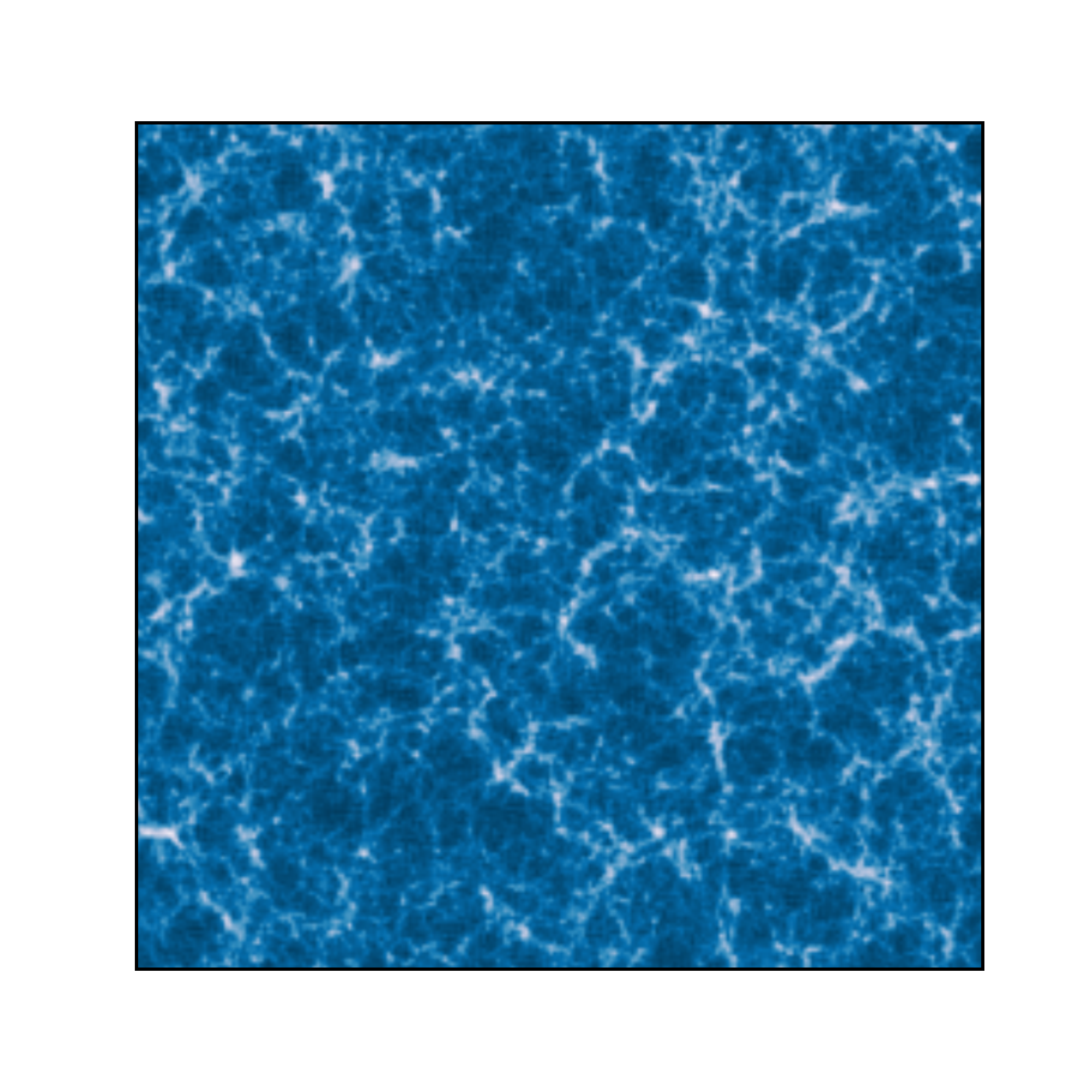}\\
\includegraphics[width=0.6\columnwidth]{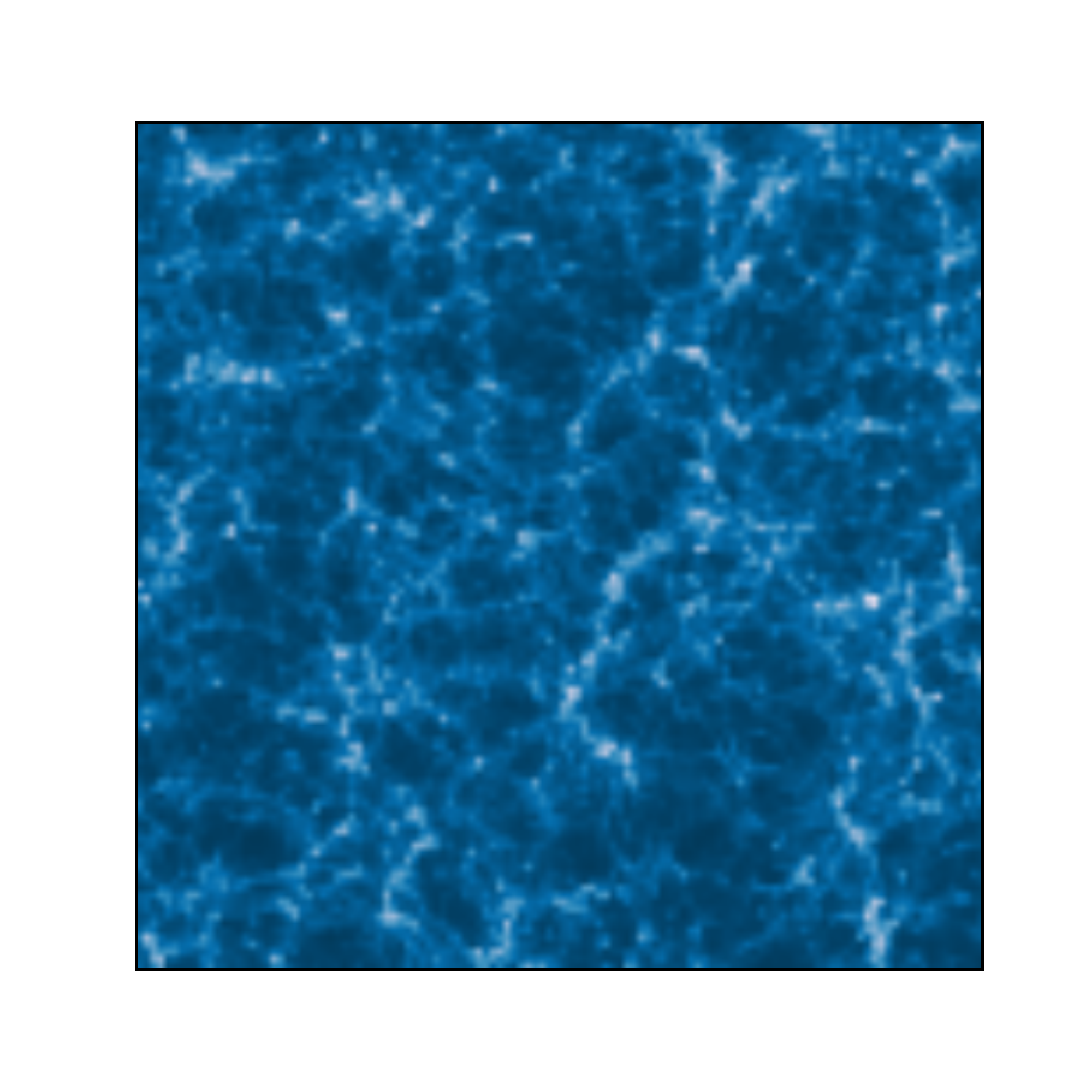}
\includegraphics[width=0.6\columnwidth]{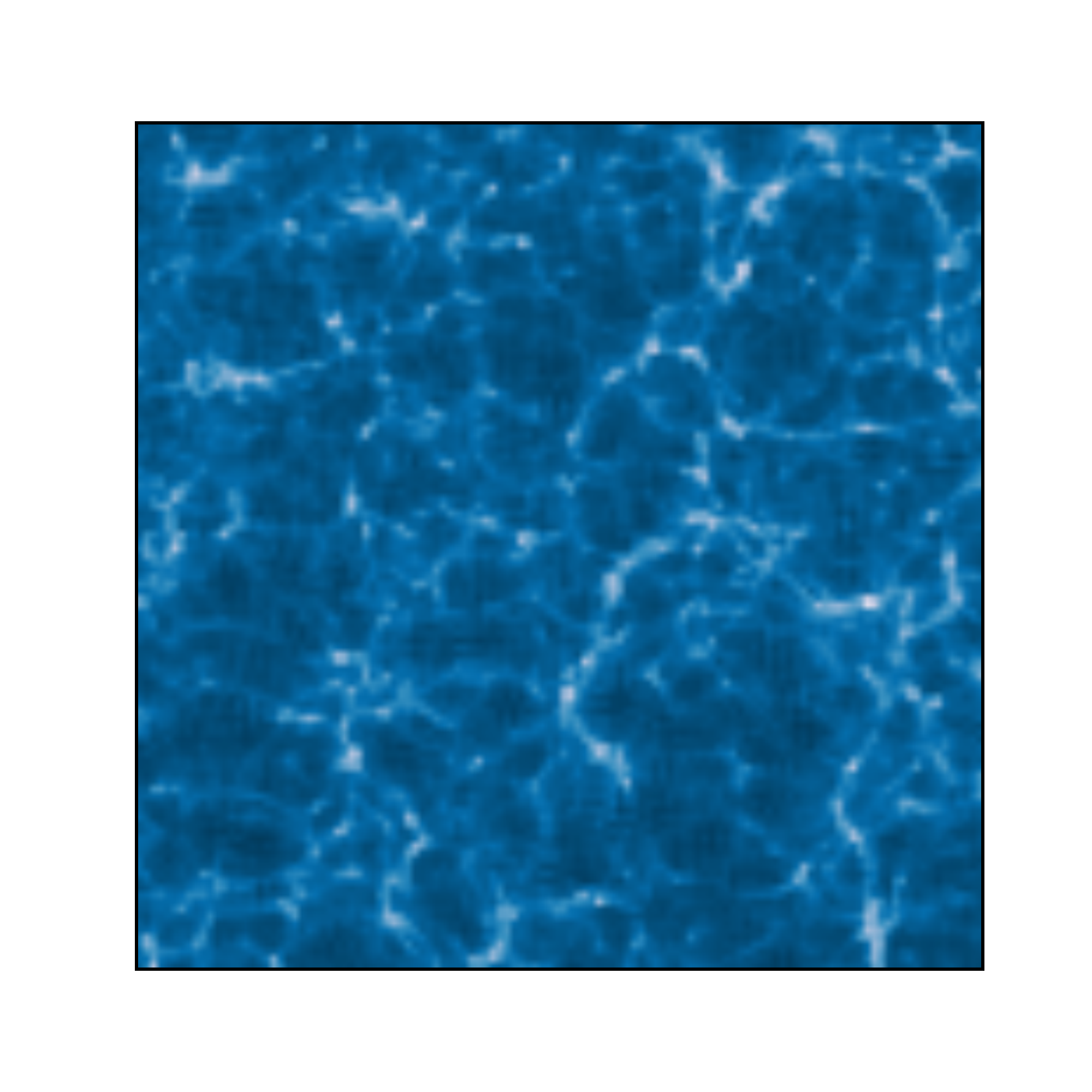}
\includegraphics[width=0.6\columnwidth]{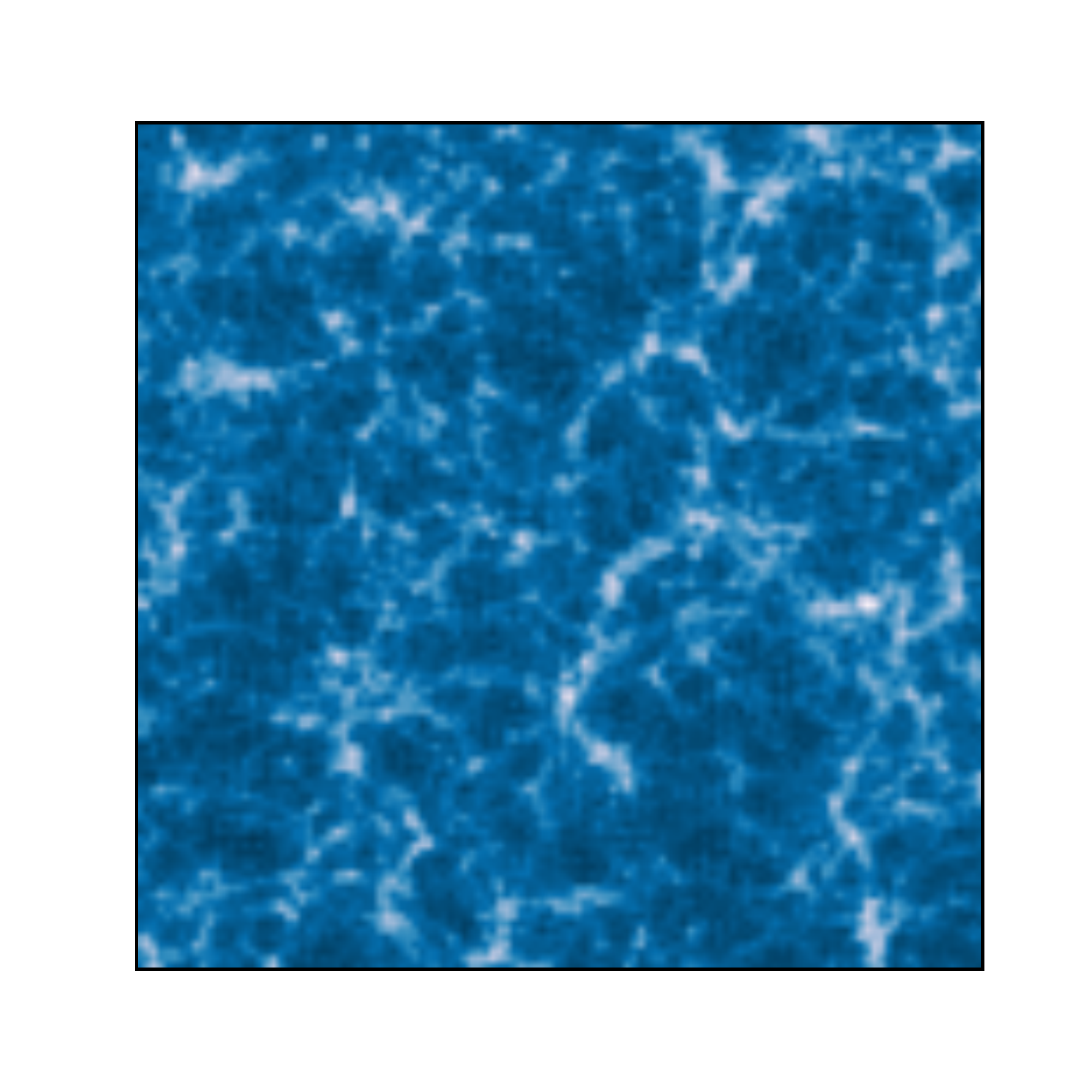}
\end{tabular}
\caption{\label{fig:darkmatter} Dark matter overdensity $\delta = \rho_{\rm m}/\rho -1$ slices of 20 $\mperh$ from the high-resolution BigMultiDark simulation (left panels) , the low-resolution \textsc{FastPM} simulation (central panels) and from the \textsc{alpt} simulation (right panels), taking a subvolume of $(1250 \;\mperh)^3$ (top panels), $(625 \;\mperh)^3$ (middle panels), and $(312.5 \;\mperh)^3$ (bottom panels). The structures in the high-resolution $N$-body simulation and the low-resolution \textsc{FastPM} simulation look very similar inspite of having very different resolutions ($3840^3$ vs $960^3$ particles). The low-resolution \textsc{alpt} simulation with a resolution of $960^3$ looks more diffuse.}
\end{figure*}

\subsection{Structure formation model}
\label{sec:sf}

Originally, the \textsc{patchy} code used Augmented Lagrangian Perturbation Theory (\textsc{alpt}, \citealt{alpt}) as a structure formation model. In this model the second order Lagrangian perturbation theory is modified by employing a spherical collapse model on small comoving scales ($r\leq 2\;\mperh$). 
Any LPT based approximation will lack the one halo term in the clustering. This can be partially compensated within the bias model, however, at the price of obtaining a less accurate description of the biasing relation.
Therefore we introduce in this work within the \textsc{patchy} code the fast particle mesh code \textsc{FastPM} \citep{fastpm}. 
In \textsc{FastPM}, the kick and drift steps of the PM codes are modified such that the linear growth of structure is exact. \citet{fastpm} demonstrates that the memory requirements of this algorithm are much lower than those of the COmoving Lagrangian Acceleration $N$-body solver \citep[\textsc{cola},][]{cola2013}. 

Moreover, \citet{fastpm} shows that running the code with relatively few time steps, and applying a friends-of-friend (hereafter fof) halo finder (\citealt{fof}) to the density field, one can accurately recover the redshift space power spectrum of the fof halos of TreePM accurate $N$-body solver (\citealt{treepm}) down to $k \sim 0.5 \; \hperm$ . The linking length of 0.2 was chosen to be consistent with other works in the literature (\citealt{cola2013}). In this work we run the \textsc{FastPM} code with 10 time steps. 

In $N$-body simulations, the dark matter density field
is evolved by solving the Boltzmann and Poisson equations
in an expanding background. Particle Mesh (PM) gravity
solvers are a class of $N$-body solvers in which the evolution
of the density field is governed by the dynamic of
dark matter particles. These dark matter particles are
evolved with a finite number of time steps.

Furthermore, in PM solvers, the density field is realized on a mesh
with an interpolation scheme (for example CIC) that assigns
particles to grid points. Then the gravitational Potential
(and subsequently the gravitational force) is estimated
by solving the Poisson equation. This grid-based
estimation of gravitational forces results in compensation
of accuracy on scales comparable to the spacing
between grid points. In \textsc{TreePM} algorithm on the contrary, small-scale
force calculation can be fully resolved by direct summation
of pairs with a spacing comparable to the spacing between
grid points. By growth of the matter overdensity
however, the computing time required for the direct summation
scheme quickly increases. 

The time steps in a PM simulation are either linearly
or logarithmically spaced in scale factor. Logarithmic
spacing of time steps have the disadvantage of losing
accuracy in terms of small-scale power and halo mass
function. On the other hand, choosing time steps that are
linear in scale factor leads to more accurate clustering
on small scales. It is important to note that finite number of time steps in PM simulations results in inaccurate
large-scale growth.

In FastPM, corrections are applied to the equations
of motions of particles such that exact large-scale growth
is imposed. In particular, the error in large scale growth is
corrected by using the Zeldovic equations of motion to modify
the kick and drift operations in a pure PM algorithm.
In a PM simulation, the drift operator changes the position
of each particle by keeping its momentum fixed. On
the other hand, the kick operator changes the momentum
of each particle while leaving its position unchanged. The
modified Kick and Drift operators of FastPM are derived
by integration of the Zeldovic Approximation equations
of motion. As a result, exact large scale growth is
imposed in FastPM and in the limit of infinitesimal time
steps the Kick and Drift operators of FastPM converge
to those of a standard PM algorithm.

In this work we will use as a reference the high-resolution $N$-body BigMultiDark simulation described in more detail in section \S \ref{sec:reference}.
A comparison of the dark matter density fields obtained with the different methods is shown in Fig.~\ref{fig:darkmatter}. While the structures in the high-resolution $N$-body simulation and the low-resolution \textsc{FastPM} simulation look very similar in spite of having very different resolutions ($3840^3$ vs $960^3$ particles), the low-resolution \textsc{alpt} simulation looks more diffuse due to the exaggerated shell crossing inherent to LPT based methods. We will study the impact of this inaccuracy in more detail in section \S \ref{sec:stats}. 

\subsection{Sampling halos from the density field}
\label{sec:bias}

In this section, we describe the statistical bias model of the \textsc{patchy} code.  This model generates halos/galaxies from a given dark matter density field and consists of: deterministic bias, stochastic bias, and an additional step for applying redshift space distortions (RSDs) to the catalogs. We describe the bias steps below and leave RSDs for a later work. 

\subsubsection{Deterministic bias}

The expected number of halos $\langle \rho_{\rm h} \rangle$ in a given volume element ${\rm d}V$ (cosmic cell) can be described in general by a deterministic bias relation $B(\rho_{\rm h}|\rho_{\rm m})$:
\be
\langle \rho_{\rm h} \rangle_{{\rm d}V} = f_{\rm h}\, B(\rho_{\rm h}|\rho_{\rm m})\,,
\ee
where $\rho_{\rm m}$ is the matter density field. The prefactor $f_{\rm h}$ is an overall normalization factor which can be determined by requiring the halo density field to have the number density of the reference sample $n_{\rm h}$, i.e., $n_{\rm h}=\langle\langle \rho_{\rm h}\rangle_{{\rm d}V}\rangle_{V}$. Formally, this can be written as 
\be
f_{\rm h} = \frac{n_{\rm h}}{\langle B(\rho_{\rm h}|\rho_{\rm m}) \rangle_{V}}\,,
\ee
where  $\langle \;.\; \rangle_{V}$ is an ensemble volume average. 
In particular, we will adopt the following compact deterministic bias model:
\ba
B(\rho_{\rm h}|\rho_{\rm m}) &=&  \underbrace{\rho_{\rm  m}^\alpha}_{\mathrm{nonlinear \; bias}} \nonumber \\ 
&\times& \underbrace{\theta\big(\rho_{\rm m} - \rho_{\rm th}\big)}_{\mathrm{threshold \; bias}} \; \times \underbrace{\exp \big(-(\rho_{\rm m}/\rho_{\epsilon})^{\epsilon}\big)}_{\mathrm{exponential \; cutoff}}\,,
\label{eq:deterministic}
\ea
where $\rho_{\rm th}$ is the density threshold which suppresses halo formation in under-dense regions, and $\alpha$ is a nonlinear bias parameter.
The threshold bias (\citealt{kaiser1984,bardeen1986,sheth2001,mo2002}) is modeled by a step function $\theta \big(\rho_{\rm m} - \rho_{\rm th}\big)$ (\citealt{kitaura2014}) and an exponential cutoff $\exp \big(-(\rho/\rho_{\epsilon})^{\epsilon}\big)$ (\citealt{neyrinck2014}). 
Therefore, for this particular bias model we have a normalization of 
\be
f_{\rm h} = \frac{n_{\rm h}}{\langle \theta(\rho_{\rm m}-\rho_{\rm th})\; \rho_{\rm m}^{\alpha}\; \exp \big(-(\rho_{\rm m}/\rho_{\epsilon})^{\epsilon}\big) \rangle_{V}}\,.
\label{eq:normalization}
\ee

The advantage of this kind of bias model is that it is flexible and it is able to incorporate additional terms and each of the terms has a physical interpretation.
The power law bias stands for one of the simplest possible nonlinear bias models: a linear Lagrangian bias in a comoving framework, which can be derived from the lognormal approximation \citep[see][]{kitaura2014}, and it resumes in one single bias parameter an infinite Taylor expansion of the dark matter density field \citep{cen1993,fry1993,delatorre}.

The threshold bias and the exponential cut-off describe the fact that halos (or galaxies) can only reside in regions which contain a minimum mass. They also represent the loss of information with respect to the full cosmic density field from selecting only gravitationally collapsed objects.

\subsubsection{Stochastic bias}

The number of halos in each cell is drawn from a Negative Binomial (NB) distribution which can be characterized by the expected number of halos in the cell $\lambda_{\rm h} = \langle \rho_{\rm h} \rangle_{{\rm d}V} \times {\rm d}V$, and a parameter $\beta$ which quantifies the stochasticity (deviation of the distribution from Poissonity) in the halo distribution. According to this model, the probability of having $N_{\rm h}$ objects in a volume element is given by
\ba
P(N_{\rm h}|\lambda_{\rm h}, \beta) &=& \underbrace{\frac{\lambda_{\rm h}^{N_{\rm h}}}{N_{\rm h}!}\, e^{-\lambda_{\rm h}}}_{\mathrm{Poisson\; distribution}} \nonumber \\ 
&\times& \underbrace{\frac{\Gamma(\beta+N_{\rm h})}{\Gamma(\beta)(\beta + \lambda_{\rm h})^{N_{\rm h}}}\times\frac{e^{\lambda_{\rm h}}}{(1+\lambda_{\rm h}/\beta)^{\beta}}}_{\mathrm{Deviation\; from\; Poissonity}}\,.
\label{eq:devpois}
\ea
For $\beta\rightarrow\infty$ we can show that the second row in the above equation goes to one. Since $\Gamma(\beta)=\frac{\Gamma(\beta+1)}{\beta}=\frac{\Gamma(\beta+N_{\rm h})}{\beta(\beta+1)\cdots(\beta+N_{\rm h}-1)}$, the first factor can be written as $\frac{\Gamma(\beta+N_{\rm h})}{\Gamma(\beta)(\beta + \lambda_{\rm h})^{N_{\rm h}}}=\frac{\beta(\beta+1)\cdots(\beta+N_{\rm h}-1)}{(\beta + \lambda_{\rm h})^{N_{\rm h}}}=\frac{(1+1/\beta)\cdots(1+(N_{\rm h}-1)/\beta)}{(1 + \lambda_{\rm h}/\beta)^{N_{\rm h}}}$. It is now straightforward to see that this goes to one for $\beta\rightarrow\infty$. The same happens for the second factor $\frac{e^{\lambda_{\rm h}}}{(1+\lambda_{\rm h}/\beta)^{\beta}}\rightarrow 1$, since $(1+\lambda_{\rm h}/\beta)^{\beta}\rightarrow e^{\lambda_{\rm h}}$ in that limit.

Given a dark matter density field $\rho_{\rm m}$, the halo density field can be constructed by drawing samples from the expected halo density field $\rho_{\rm h}$ with the Negative-Binomial (hereafter NB) distribution (Eq.~\ref{eq:devpois}). This is inspired by the fact that the excess probability of finding halos in high density regions generates over-dispersion \citep[][]{somerville2001,miranda2002}. This over-dispersion is modeled by a NB distribution (\citealt{kitaura2014,neyrinck2014}). 

The stochastic bias stands for the shot noise from the transition of the continuous dark matter field to the discrete halo (or galaxy) distribution. As predicted by \citet{Peebles1980}, it produces a dispersion larger than Poisson, as long as the two-point correlation function remains positive below the scale of the cell size. This is captured by the negative binomial PDF (Eq.~\ref{eq:devpois}).

\subsection{Constraining the bias model}
\label{sec:mcmc}

Production of approximate mock catalogs with \textsc{patchy} requires a reference catalog constructed from the observations or based on an accurate $N$-body simulation. 
We aim at constraining the parameters describing the deterministic bias $\{\delta_{\rm th},\alpha, \rho_{\epsilon},\epsilon\}$, and the parameter that governs the stochasticity of the halo population $\{\beta\}$.

The bias parameters are estimated such that the statistical summaries of the halos (galaxies) in the \textsc{patchy} mocks match the statistical summaries of the halos (galaxies) in the reference catalog. The set of statistical summaries of the catalog can in principle include number density, bivariate probability distribution function or number of counts-in-cells $\rho$ (), two-point statistics $\xi_{2}$, and higher-order statistics such as the three-point statistics $\xi_{3}$. Note that the number of counts-in-cells is defined as the number of cells that contain a given number of halos.

By construction, the \textsc{patchy} mocks reproduce the exact number density of objects in the reference catalog. This comes from the particular choice of normalization in the deterministic bias relation (see Eqs.~\ref{eq:deterministic},\ref{eq:normalization}). In this work, we follow \citet{kitaura2015} and constrain the bias parameters with the halo PDF and the two-point statistics $\xi_{2}$. These two quantities can be computed very fast and the skewness of the halo PDF determines the three point statistics. Given the bias parameters found by fitting the PDF and the two-point statistics, we will demonstrate a comparison between the approximate mocks and the reference catalog in terms of the two- and three-point statistics. 

We simultaneously fit the real-space power spectrum $P(k)$ and the PDF $\rho(n)$ of the \textsc{patchy} halo density field to $P(k)$ and $\rho(n)$ measured for the BigMuliDark halo catalog. Specifically, constraints on $\theta = \{\delta_{\rm th},\alpha, \rho_{\epsilon},\epsilon, \beta\}$ are found by sampling from the posterior probability $p(\theta|\mathrm{data}) \propto p(\mathrm{ref}|\theta)p(\theta)$, where $\mathrm{ref}$ denotes the combination $\{P_{\rm ref} (k), \rho_{\rm ref}(n)\}$, and the likelihood $p(\mathrm{ref}|\theta)$ is given by

\be
p\big(\mathrm{ref}|\theta \big) = p\big(P_{\rm ref}(k)|\theta \big ) p\big(\rho_{\rm ref}(n)|\theta \big),
\label{eq:probs}
\ee
where Gaussian likelihoods are assumed for $P_{\rm ref}(k)$ and $\rho_{\rm ref}(n)$:
\ba
p\big(P_{\rm ref}(k)|\theta \big ) &=& \prod_{k} \frac{1}{\sqrt{2\pi\sigma_k^2}} \nonumber \\ 
&\times& \exp\Big[\frac{-\big(P_{\rm ref}(k)-P_{\rm mock}(k)\big)^2}{2\sigma_k^{2}}\Big],\\
p\big(\rho_{\rm ref}(n)|\theta \big ) &=& \prod_{n} \frac{1}{\sqrt{2\pi\sigma_n^2}} \nonumber \\ 
&\times& \exp\Big[\frac{-\big(\rho_{\rm ref}(n)-\rho_{\rm mock}(n)\big)^2}{2\sigma_n^{2}}\Big].
\ea

As a result:
\ba
-2\ln p(\mathrm{ref}|\theta) &=& \sum_{k}\Big[\frac{\big(P_{\rm ref}(k)-P_{\rm mock}(k)\big)^{2}}{\sigma^{2}_{k}}  \nonumber \\
&+& \ln(2\pi\sigma_k^2) \Big] \nonumber \\
&+& \sum_{n}\Big[\frac{\big(\rho_{\rm ref}(n)-\rho_{\rm mock}(n)\big)^{2}}{\sigma^{2}_{n}}  \nonumber \\
&+& \ln(2\pi\sigma_n^2)\Big] \label{eq:like}
\ea

For the purpose of estimating the bias parameters, we find it sufficient to assume simple uncorrelated noise terms $\{\sigma_{k},\sigma_{n}\}$ in the above likelihood (\ref{eq:like}). We assume $\sigma_{k}^{2}$ to be $4\pi^{2} P_{\rm ref}^{2}(k)/(V_{\rm box}k^{2}\Delta k)$, and $\sigma_n^{2}$ to be $N_n$ where $N_n$ is the number of cells containing $n$ number of halos (including parent halos and subhalos).

In Eq.~\ref{eq:like}, a Gaussian likelihood is assumed. It is worth 
noting that the relative weight between the two terms in the likelihood function is determined by their corresponding bin sizes. For instance, increasing the size of the $n$th bin in the term corresponding to the PDF in Eq.~\ref{eq:like}, will increase $N_n$ and as a result $\sigma_n$. This gives more weight to the power spectrum and reduces the prediction power of the halo PDF. We find that giving more weight to the power spectrum results in higher deviations from Poissonity (higher stochasticity). Higher stochasticity (lower $\beta$) leads to enhancement of small-scale power but at the cost of reducing the quality of fit for the halo PDF. This leads to less accurate bispectrum.

Furthermore, we choose a flat prior for all parameters of the bias model with the following lower and upper bounds: $-1<\delta_{\rm th}<2$, $0<\alpha<1$, $0<\beta<1$, $0<\rho_{\epsilon}<1$, and $0<\epsilon<1$.

For sampling from the posterior probability, given the likelihood function (Eq.~\ref{eq:like}) and the prior, we use the affine-invariant ensemble MCMC sampler (\citealt{goodmanweare}) and its implementation \textsc{emcee} (\citealt{emcee}). In particular, we run the \textsc{emcee} code with 10 walkers and we run the chains for at least 2000 iterations. We discard the first 500 chains as burn-in samples and use the remainder of the chains as production MCMC chains. Furthermore, we perform Gelman-Rubin convergence test (\citealt{grtest}) to ensure that the MCMC chains have reached convergence.

\subsection{Comparison with other approximate methods}

Pioneering fast halo/galaxy generating methods have relied on approximate gravity solvers based on Lagrangian perturbation theory (LPT: \citealt{buchert1993,bouchet1995,catelan1995,scocci2002}) to compute the positions and masses of the objects, such as \textsc{Pinocchio} (Zeldovich: \citealt{monaco2002,monaco2013}, 3LPT: \citealt{monaco2016}), and \textsc{PThalos} (2LPT: \citealt{pthalo,manera2015}).

This has the disadvantage of being affected by an inaccurate description of the small scale clustering, and, in particular, of missing the one halo term contribution. As a consequence, the power spectra of such catalogs have systematic deviations towards high values of $k$, already deviating about 10$\%$ at $k\sim0.2\; \hperm$ \citep[][]{monaco2013}.

While fast particle mesh solvers, such as \textsc{cola} or \textsc{FastPM}, are much more precise than LPT based approaches, they are  still computationally too expensive to be suitable for massive production, if one is trying to resolve all the necessary structures required to model next generation of galaxy surveys.


Therefore four methods were recently proposed: \textsc{patchy} \citep{kitaura2014}, \textsc{QPM} \citep{qpm}, \textsc{EZmocks} \citep{eazymock}, and \textsc{HALOGEN} \citep{halogen}, which do not try to resolve halos (nor galaxies) with the approximate gravity solvers, but just get a reliable large scale dark matter field, which can then be populated with some bias prescription. The gravity solver thus only needs to be accurate on a certain scale, then the halo/galaxy-dark matter connection is exploited to reach a high accuracy, as described above.

These methods use both different gravity solvers and different bias models. While \textsc{patchy} originally relies on \textsc{alpt}, \textsc{QPM} uses a quick particle mesh solver, \textsc{EZmocks} uses the Zeldovich linear LPT, and \textsc{HALOGEN} uses 2LPT.
But more importantly the bias prescriptions follow very different philosophies. \textsc{QPM} uses a rank ordering scheme relating the halo mass to density peaks. Large scale halo bias and halo mass function are recovered in this algorithm. Similarly, \textsc{HALOGEN} relies on first, sampling a number of halo masses from a mass function in discrete mass bins. Then in each mass bin (rank-ordered from high mass to low mass), halos are assigned to cosmic cells according to a distribution function governed by a parameter. This parameter is determined by fitting the two-point function of halos in that mass bin to that of an accurate reference $N$-body catalog. Therefore, this algorithm is designed to recover the mass-dependent halo bias of a target $N$-body simulation. However, a recent study  demonstrated  that the dependence of the halo mass to its environment is not trivial \citep[see][]{zhao2015}.  \textsc{EZmocks}, on the other hand, first modifies the initial power spectrum introducing a tilt to adjust the final two point statistics, correcting hereby the missing one halo term of the approximate gravity solver. Second it imposes the halo PDF, which was shown to determine the 3pt statistics \citep{kitaura2015}.

\textsc{patchy} on the other hand follows a more physical approach, relying on an effective analytical stochastic bias model. 
In this sense the statistics is not directly imposed as in \textsc{EZmocks}, but fitted through the bias parameters. In fact \textsc{patchy} was shown to be considerably more accurate than \textsc{EZmocks} when assigning halo masses \citep{zhao2015}, and than \textsc{QPM} when fitting the two and three point statistics of the luminous red galaxies (LRGs) in the Baryon Oscillation Spectroscopic Survey (BOSS)  \citep{kitaura2016}. 

Relying only on particle mesh gravity solvers for production of mock catalogs is more computationally demanding. That is, in order to resolve all structures (and substructures) one needs to run a PM code with a higher resolution than what is required by \textsc{patchy}. For instance \citet{chuang2015} found that in order to reproduce distinct (parent) halos of the BigMultiDark simulation, \textsc{cola} needs a particle-mesh size of $1280^3$. This is far away from reproducing the substructures (subhalos). \textsc{patchy} on the other hand, requires a smaller grid size of $960^3$ for production of distinct (parent) halos and subhalos \citep[see][]{chuang2015}. 
In this work we use a grid size of $960^3$ for generation of both \textsc{alpt} and \textsc{FastPM} density fields.

Furthermore, the findings of \citet{fastpm} suggest that in order to recover the halo mass function of an accurate $N$-body simulation with a PM solver and a friends-of-friends halo finding algorithm, one needs to choose a force resolution (the ratio of the grid size and the number of particles on one side of a PM simulation) greater than or equal to two. That is, the mass resolution of a PM simulation for which the grid size and the number of particles are equal, is not sufficient to resolve distinct halos. On the other hand, using simulations with low mass resolution and stochastic biasing methods such as \textsc{patchy} and \textsc{QPM} can accurately recover the clustering statistics of halos and subhalos in accurate $N$-body simulations.

Moreover, the approach we follow in \textsc{patchy} tests the validity range of effective bias prescriptions commonly used in large scale structure analysis methods \citep[see e.g.][]{ata2015}. Now for the first time we include a robust MCMC sampling scheme to determine the bias parameters, and have improved the gravity solver with \textsc{FastPM}.

\begin{figure*}
\includegraphics[width=2\columnwidth]{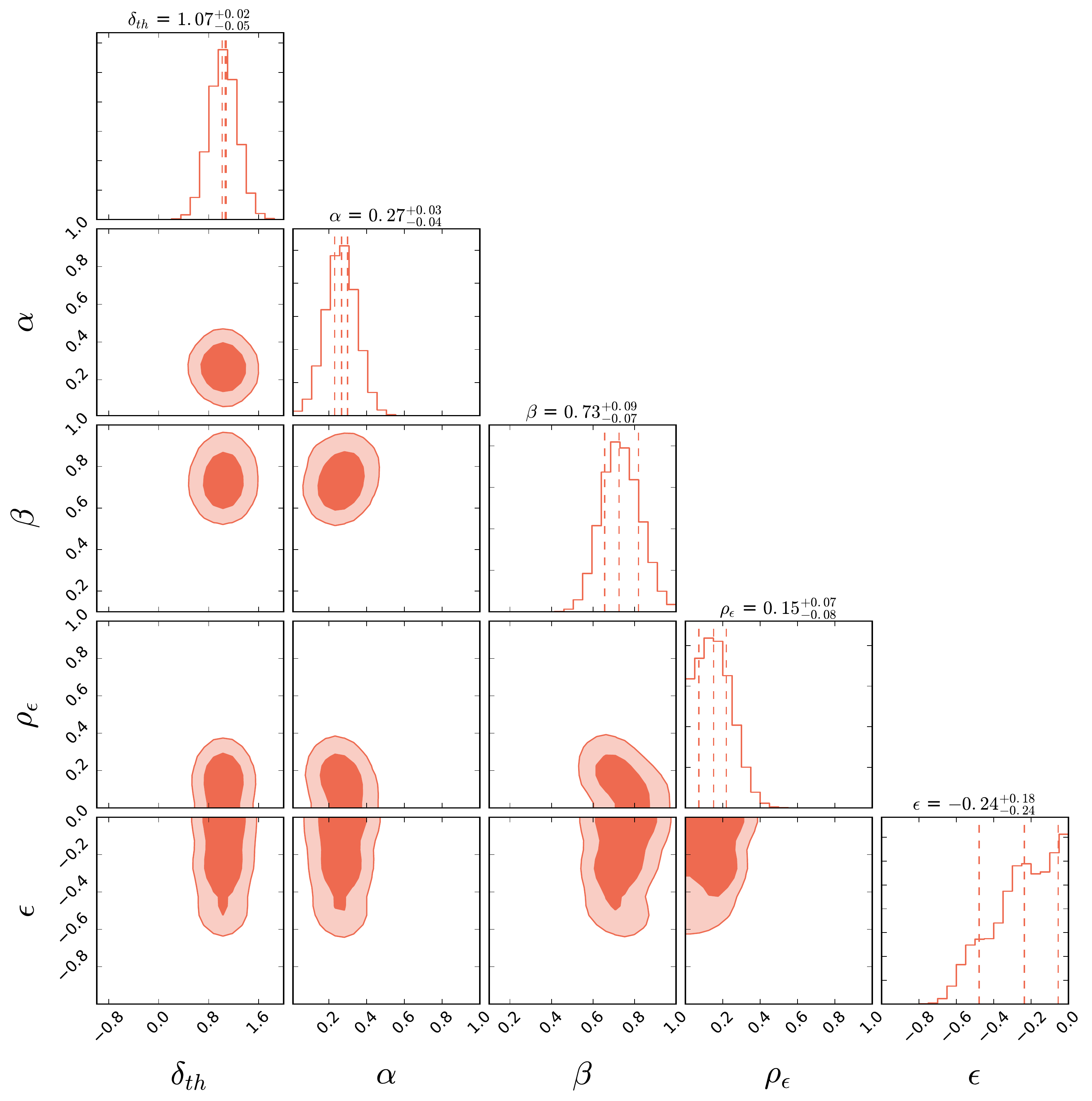}
\caption{\label{fig:bias} Posterior probability distribution of the \textsc{patchy} bias parameters $\{\delta_{\rm th},\alpha,\beta,\rho_{\epsilon},\epsilon\}$. The contours mark the 68$\%$ and the 95$\%$ confidence intervals of the posterior probabilities. The vertical lines represent the best estimate and the error bars corresponding to the 50$\%$ and 68$\%$ confidence intervals obtained from the marginalized posterior distribution over \textsc{patchy} bias parameters. This plot is made using the open-source software \textsc{corner} (\citealt{corner}).}
\end{figure*}

\begin{figure*}
 \begin{tabular}{cc}
\includegraphics[width=\columnwidth]{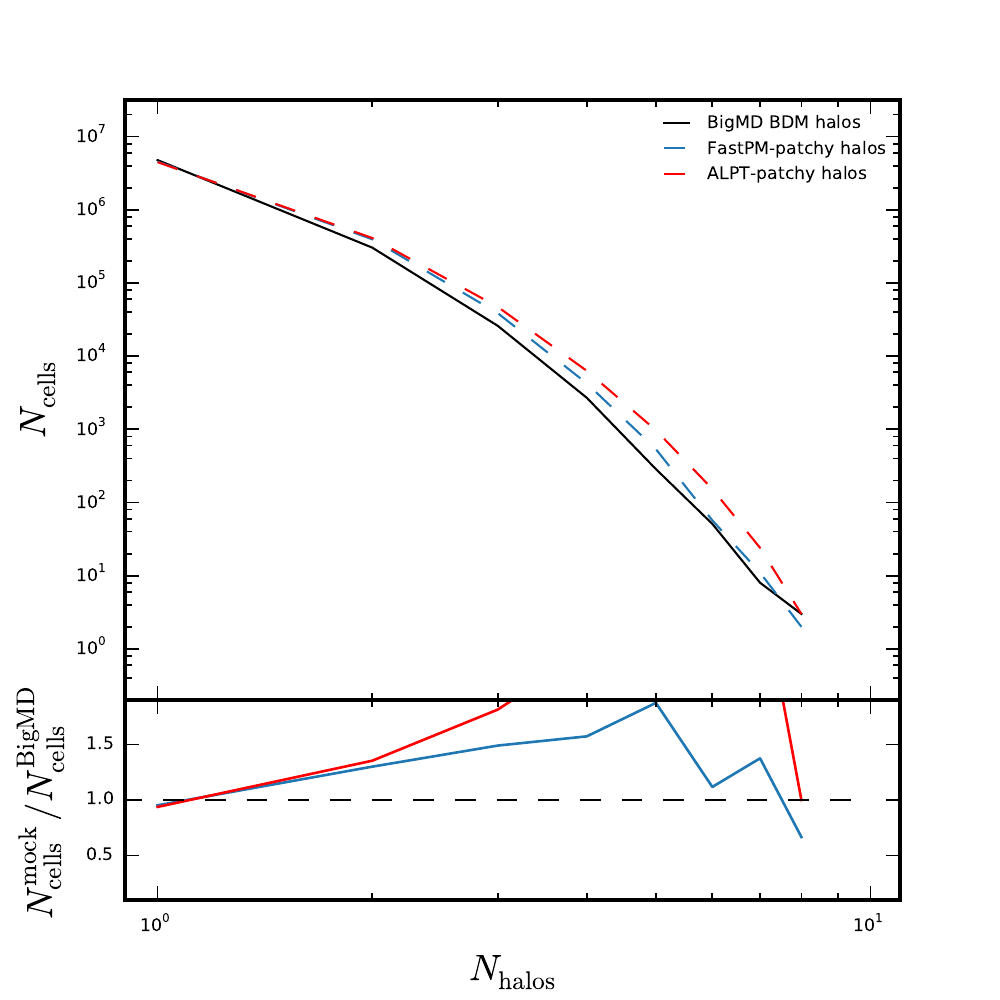}
\includegraphics[width=\columnwidth]{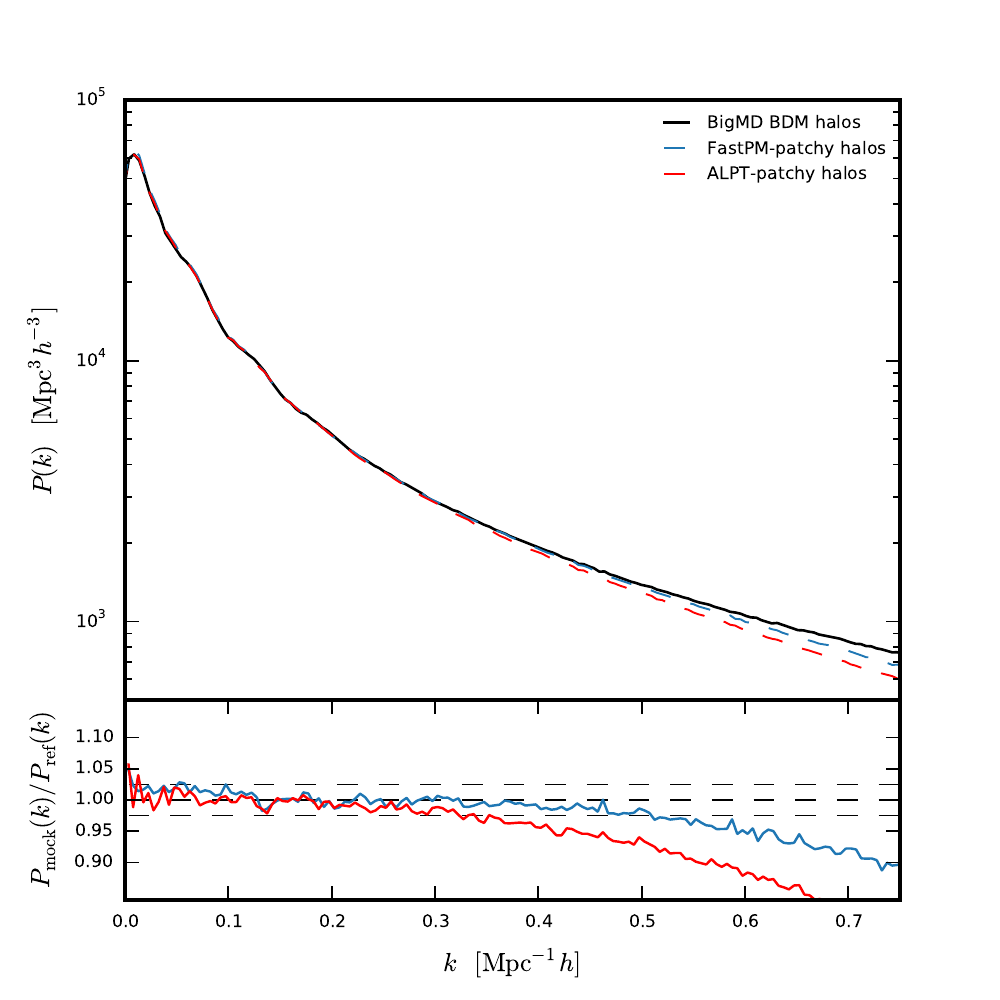}
\end{tabular}
\caption{\label{fig:pdfpower} Top: Demonstration of the halo bivariate probability distribution function of halos (halo counts-in-cells) in the BigMultiDark simulation (shown in black) and in the \textsc{FastPM}-\textsc{patchy} simulation (shown in blue) and in the \textsc{alpt}-\textsc{patchy} simulation (shown in red) on the left. Comparison between the real-space power spectrum of the BDM halos (shown in black) in the reference BigMultiDark simulation and that of the halos in the \textsc{FastPM}-\textsc{patchy} (\textsc{alpt}-\textsc{patchy}) simulation shown in blue (red) on the right. 
Bottom: Ratio between the halo PDFs of the approximate mocks and halo PDF of the BigMultiDark simulation on the left. Ratio between the halo power spectra of the approximate mocks and the halo power spectrum of the BigMultiDark simulation on the right.}
\end{figure*}

\begin{figure*}
\vspace{-0.5cm}
\begin{tabular}{cc}
\hspace{-0.4cm}
\includegraphics[width=1.1\columnwidth]{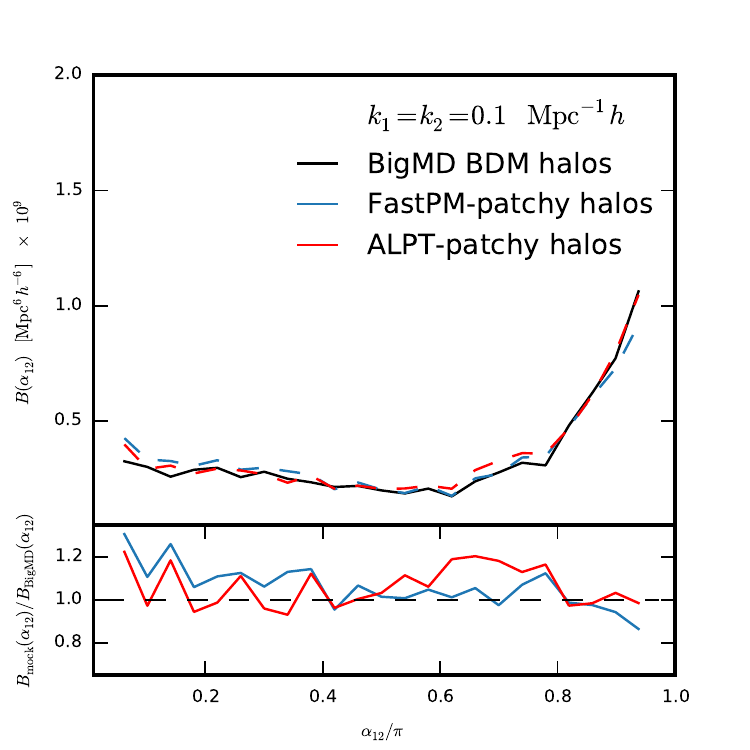}
\includegraphics[width=1.1\columnwidth]{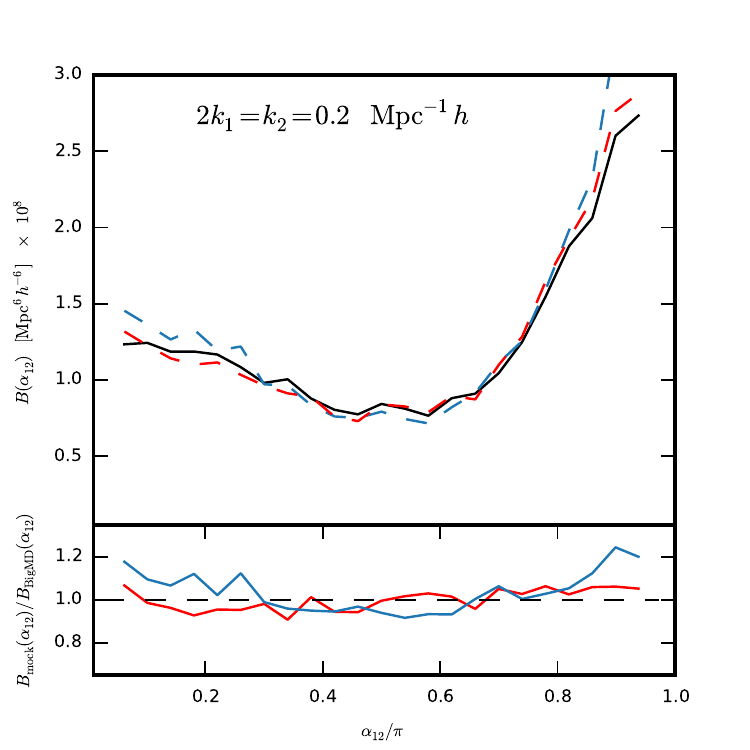}
\vspace{-0.93cm}
\\
\hspace{-0.4cm}
\includegraphics[width=1.1\columnwidth]{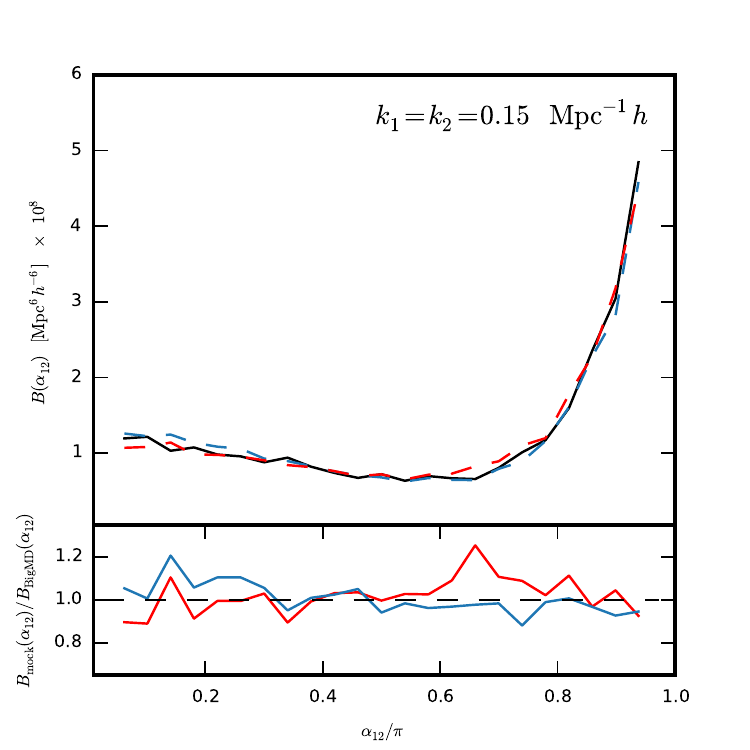}
\includegraphics[width=1.1\columnwidth]{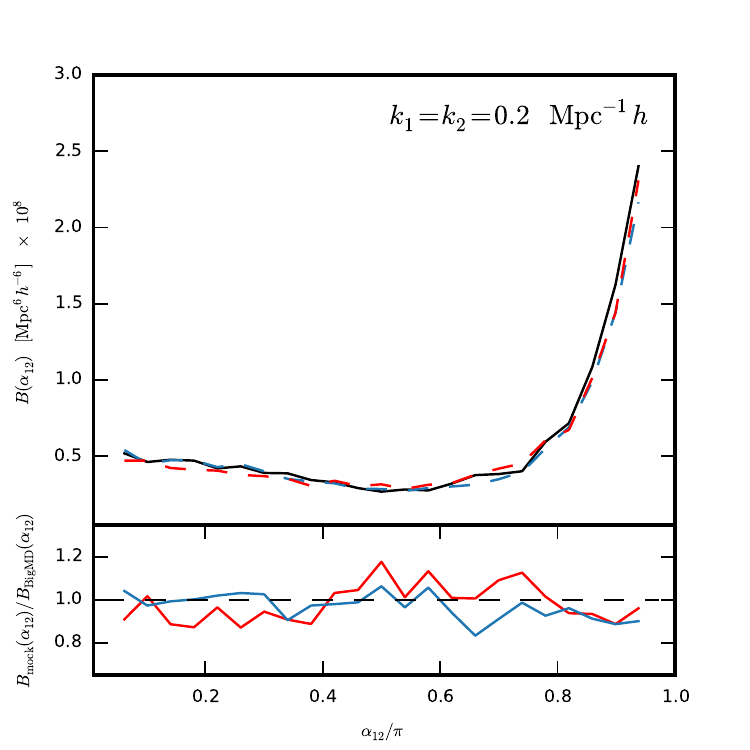}
\vspace{-0.93cm}
\\
\hspace{-0.4cm}
\includegraphics[width=1.1\columnwidth]{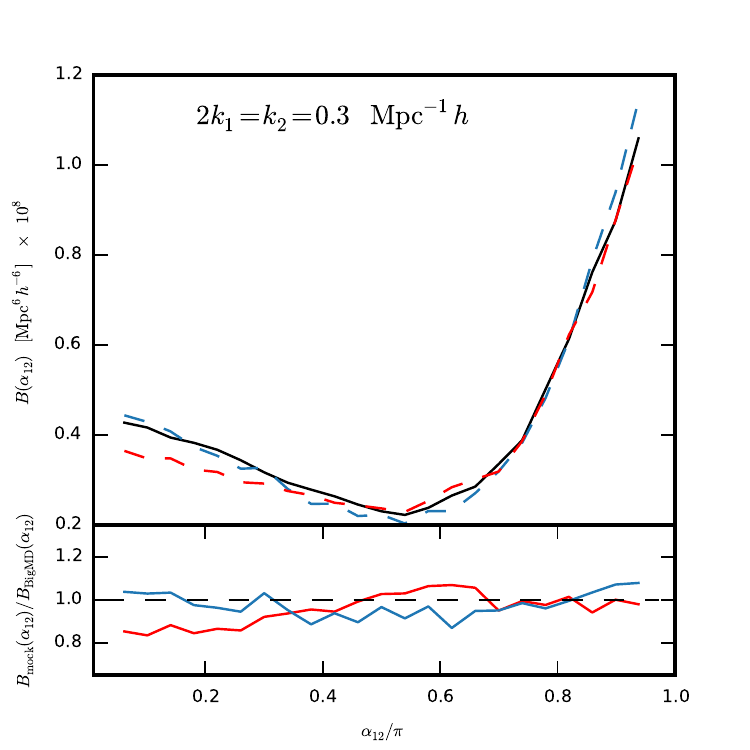}
\includegraphics[width=1.1\columnwidth]{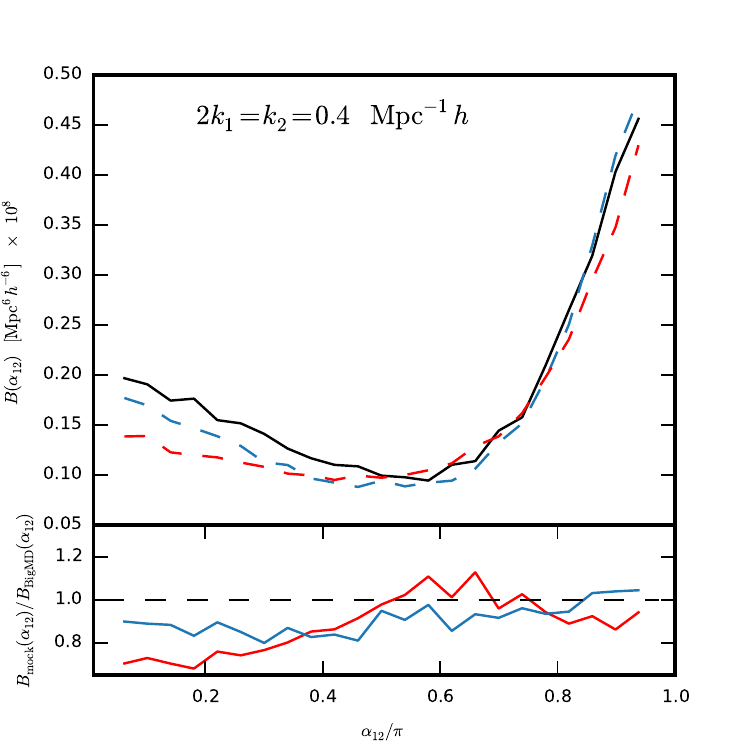}
\vspace{-0.25cm}
\end{tabular}
\caption{\label{fig:bispec} Real-space bispectrum of the BigMD BDM halos and that of the approximate mocks as a function of angle $\alpha_{12}$ between $\mathbf{k}_1$ and $\mathbf{k}_{2}$ for $k_{1}=k_{2}=0.1\; \hperm$ (upper left), $2k_{1}=k_{2}=0.2\; \hperm$ (upper right), $k_{1}=k_{2}=0.15\; \hperm$ (middle left), $k_{1}=k_{2}=0.2\; \hperm$ (middle right), $2k_{1}=k_{2}=0.3\; \hperm$ (lower left), and $2k_{1}=k_{2}=0.4\; \hperm$ (lower right). The BigMD is represented by the solid black line, while \textsc{alpt}-\textsc{patchy} is represented by the dashed red line, and \textsc{FastPM}-\textsc{patchy} is represented by the dashed blue line.}
\end{figure*}

\section{Code Integration}\label{sec:code_integration}

In this section, we briefly explain how different pieces of the \textsc{patchy} code are connected. The first step is specification of the cosmological parameters and the initial matter power spectrum. Afterwards, the initial conditions can be provided to the \textsc{FastPM} code in the form of a white noise (compatible with the initial power spectrum), or they can be randomly drawn from the power spectrum in the \textsc{FastPM} code.

Once the dark matter particles are evolved with the
particle-mesh code until the final redshift, the positions
of particles are recorded with the \textsc{FastPM} code. Then
within the \textsc{patchy} code, the final particle positions are
painted onto a mesh with Cloud-in-Cell (CIC) algorithm.
A set of \textsc{patchy} bias parameters and the density mesh
are the main ingredients for generating a catalog of tracers
(galaxies/halos) of the dark matter density field.

The bias parameters are estimated by MCMC in the following way: 
In an MCMC wrapper, the posterior probability distribution is defined as a function of \textsc{patchy} bias parameters. We use the \textsc{emcee} code to sample from this posterior probability distribution in the wrapper. For each set of bias parameters, a catalog of galaxies/halos is generated from the density mesh with the \textsc{patchy} code. Once a catalog is generated, the PDF and the real-space power spectrum of halos in the mock catalogs are evaluated for computation of the likelihood~(\ref{eq:like}).

\section{Demonstration on an accurate $N$-body based halo catalog}\label{sec:test}

In this section we present the application of the above described method to a well studied case: the halo distribution required to describe the CMASS LRG sample of the BOSS survey \citep{white2011,Dawson2013}.
First, we briefly describe the reference catalog and then present a detailed statistical analysis of the results.

\subsection{Reference catalog}
\label{sec:reference}

For the reference simulation used in this work we rely on the Bound-Density-Maxima (BDM, \citealt{bdm}) halo catalogs in the $z=0.5618$ snapshot of the BigMultiDark-Planck high 
resolution $N$-body simulation \citep{multidark}. This simulation was carried out using the L-Gadget2 code \citep{gadget}, following the Planck $\Lambda$CDM cosmological parameters 
$\Omega_{\rm m} = 0.307$, $\Omega_{\rm b} = 0.048$, $\Omega_{\Lambda} = 0.693$, $\sigma_{8} = 0.823$, $n_{\rm s}=0.96$, 
$h=0.678$. The box size for this $N$-body simulation is 2500 $\mperh$, the number of simulation particles is 3840$^3$, 
the mass per simulation particle $m_{\rm p}$ is $2.359 \times 10^{10} \; h^{-1} M_{\odot}$, and the gravitational softening length 
$\epsilon$ is 30 $h^{-1} \rm{kpc}$ at high-$z$ and 10 $h^{-1} \rm{kpc}$ at low-$z$.

A minimum mass cut of $0.5 \times 10^{13} \; h^{-1} M_{\odot}$ has been applied to the halo catalog so that it matches with the number density of the SDSS III-BOSS CMASS galaxy catalog \citep{white2011,Dawson2013}. After applying the mass cut, the number density of the final catalog is 3.5 $\times 10^{-4}$ $(\hperm)^{3}$. The MultiDark-\textsc{patchy} galaxy catalogs \citep{kitaura2016} are calibrated against BOSS-HAM catalogs which were constructed by populating the halos in different snapshots of the BigMultiDark simulation using halo abundance matching \citep{sergio2016}.   

Evaluation of $P(k)$ and $\rho(n)$ for a set of bias parameters requires running the forward model of generating halos from the matter density field. Therefore, in order to speed up the fitting procedure we run the \textsc{patchy} code with a smaller box size of 625 $\mperh$ and grid size of 240 in each dimension. This choice of box and grid size preserves the resolution. Furthermore, running the \textsc{patchy} code and computing the statistics of the halo catalogs in a smaller box size significantly reduces the computational time needed for constraining the bias parameters.

\subsection{Bias parameters}

The first step in our pipeline consists of producing the large scale dark matter field on a mesh. We use the down-sampled  white noise of the BigMultiDark simulation from $3840^3$ to $960^3$ cells to estimate the initial conditions used for both \textsc{FastPM} and \textsc{alpt} runs, as shown in Fig.~\ref{fig:darkmatter}. The dark matter particles are then assigned to a mesh of $960^3$ cells with clouds-in-cells (CIC), which we define as the large scale dark matter density field $\rho_{\rm m}$ required for Eqs.~\ref{eq:deterministic}, \ref{eq:normalization}, \ref{eq:devpois}. 

After running the MCMC chains with the method described in section \S \ref{sec:method}, we find constraints on the bias parameters of such equations. These constraints are summarized in Fig.~\ref{fig:bias}.  
The threshold bias parameter $\delta_{\rm th}$ is found to be 1.07 which is equivalent to sampling halos from the regions of high matter overdensity. This supports our intuition that massive halos are generated from high density regions. Our estimated value of the nonlinear bias parameter $\alpha$ is $\sim$ 0.2. These values are qualitatively consistent with \textsc{alpt} \citep{kitaura2014}, although the threshold bias is slightly reduced and the power law bias is slightly higher (parameters with \textsc{alpt}: $\delta_{\rm th} \sim 1.2$ and $\alpha \sim 0.12$).

The parameter that governs the deviation from Poissonity $\beta$ is found to be $~ 0.73$. This value is significantly larger than the one found with \textsc{alpt} (about 0.6), i.e., indicating that the deviation from Poissonity is not so pronounced, as previously found. The reason for this, is that Lagrangian perturbation theory does not manage to model the one halo term, as done with \textsc{FastPM}. Therefore a larger deviation of Poissonity had to be assumed to fit the power spectrum towards small scales, as is demonstrated here. In this sense, a more accurate description of the large scale dark matter field permits us to reduce the stochasticity in the halo distribution.

Furthermore, parameters corresponding to the exponential cutoff term in the deterministic bias relation $\{\rho_{\epsilon},\epsilon\}$ are estimated to be $\sim \{0.15,-0.24\}$. While the constraints on both parameters of the exponential cutoff bias are consistent with zero, their presence, albeit being small, is essential in a more accurate modeling of the halo bivariate PDF and the halo bispectrum. By including these extra parameters we demonstrate the flexibility and  efficiency of the code to incorporate complex bias models. Furthermore, we believe that the  exponential cutoff term will become crucial when considering smaller mass halos, which have a non negligible probability of residing in low density regions \citep[][]{neyrinck2014}.

\subsection{Statistical comparison}
\label{sec:stats}

In this section we discuss the statistical comparisons between the BDM halo catalog of the BigMultiDark simulation and the halo catalog generated from our method. In particular, the \textsc{FastPM}-\textsc{patchy} mock is generated using the best-fit bias parameters (see Fig.~\ref{fig:bias}). For the \textsc{alpt}-\textsc{patchy} mocks we rely on the parameters found from previous \textsc{patchy} studies \citep{kitaura2016}. The halo statistical summaries presented in this work are the number density, the bivariate halo probability distribution function (halo counts-in-cells), the real-space power spectrum and the real-space bispectrum. 

By construction our method reproduces the exact number density of halos in the reference catalog (Eq.~\ref{eq:normalization}). We observe that the bivariate PDF (or halo counts-in-cells) of the reference catalog can be reproduced with good accuracy (Fig.~\ref{fig:pdfpower}). 

In terms of the agreement between halo PDF of approximate mock catalog and that of the BigMultiDark simulation, we find that significant improvement can be achieved when halos are sampled from the \textsc{FastPM} dark matter density field. 

Furthermore, we present our comparison in terms of the power spectrum $P$ and the bispectrum $B$ which are the two-point function and the three-point function in Fourier space. Given the Fourier transform of the halo density field $\delta_{h}(\mathbf{k})$, the power spectrum and the bispectrum are defined as follows
\ba
\langle \delta_{h}(\mathbf{k}_{1}) \delta_{h}(\mathbf{k}_{2})\rangle &=& (2\pi)^{3} P(k_{1}) \delta^{D}(\mathbf{k}_{1}+\mathbf{k}_{2}), \\
\langle \delta_{h}(\mathbf{k}_{1}) \delta_{h}(\mathbf{k}_{2}) \delta_{h}(\mathbf{k}_{3})\rangle &=& (2\pi)^{3} B(\mathbf{k}_{1},\mathbf{k}_{2}) \delta^{D}(\mathbf{k}_{1}+\mathbf{k}_{2}+\mathbf{k}_{3}), \nonumber \\
\ea
where $\delta^{D}$ is the Dirac delta function. The shot-noise contribution to the power spectrum and bispectrum is modeled in the following way:
\ba
P_{\mathrm{sn}}(k) &=& \frac{1}{\bar{n}}, \\
B_{\mathrm{sn}}(\mathbf{k}_{1},\mathbf{k}_{2}) &=& \frac{1}{\bar{n}} [P(k_{1}) + P(k_{2}) + P(k_{3})] + \frac{1}{\bar{n}^{2}},
\ea
where $\bar{n}$ is the halo number density and $k_{3}=|\mathbf{k1}+\mathbf{k2}|$.

Our methodology is able to reproduce the halo power spectrum of the reference with $\sim 2.5\%$ accuracy to $k \sim 0.4 \; \hperm$ (within 5\% up to $k \sim 0.6 \; \hperm$) which corresponds to  nonlinear regimes (Fig.~\ref{fig:pdfpower}). We have also run our method ignoring the PDF in the posterior sampling, yielding accurate power spectra up to $k\sim 1 \; \hperm$. \citet[][]{kitaura2014} also reported accurate power spectra up to high $k$, however, using an arbitrary threshold bias of zero. In a later work additionally fitting the PDF, it was found that the power spectra are accurate within 2\% up to $k\sim 0.3 \; \hperm$ \citep[][]{kitaura2015}, in agreement with what is found here using \textsc{alpt}. 
An even higher accuracy will require a more  complex bias model and a proper modeling of the clustering on sub-Mpc scales, differentiating between centrals and satellites. The current version of \textsc{patchy} randomly assigns dark matter particle positions to halos sampled in a given cell.
The bias model could be augmented with nonlocal bias terms following \citet[][]{mcdonald2009}. We have neglected in this study the perturbation bias term used in \citet{kitaura2016} (in an attempt to compensate for the missing power towards small scales), where the limit in the $\sim$ 2\% level accuracy was found to be around $k\sim 0.3 \; \hperm$. Omitting the perturbation theory term also allows for a fair comparison with the study presented in \citet[][]{kitaura2015} and is not necessary when using \textsc{FastPM}. 



Fig.~\ref{fig:pdfpower} showed an improved PDF when relying on \textsc{FastPM}. This is expected to have an impact in the three point statistics, which in fact yields better fits towards small scales, as we discuss below. 
We show our results in terms of bispectrum for six different values of $|\mathbf{k}_1|$ and $|\mathbf{k}_2|$ as a function of the angle between the two vectors $\alpha_{12} = \angle (\mathbf{k}_1 , \mathbf{k}_2)$. The adopted wave numbers are $k_1=k_2=0.1,\; 2k_1=k_2=0.2,\; k_1=k_2=0.15,\; k_1=k_2=0.2,\; 2k_1=k_2=0.3,\; 2k_1=k_2=0.4$ (all wave numbers are expressed in units of $\hperm$). 

We find that in general for both \textsc{alpt} and \textsc{FastPM} there is good agreement between the bispectrum measured from our approximate mock catalogs and that of the BigMultiDark simulation (Fig.~\ref{fig:bispec}). 
Deviations as large as 15-20\% are expected, as we are using a down-sampled  white noise of the BigMultiDark simulation from $3840^3$ to $960^3$ cells and are on the level of what was found in \citet[][]{kitaura2015}. 


For configurations corresponding to smaller scales ($2k_1=k_2=0.3\;\hperm$, $2k_1=k_2=0.4\;\hperm$), the agreement between the bispectra of our approximate mock halo catalogs and the BigMultiDark halos improves when we sample halos from the \textsc{FastPM} density field. This improvement is dramatic when compared to \textsc{EZmocks} \citep[see real-space lines in the lower panels in Fig.~5 of][]{eazymock}. 

\citet{chuang2015} presents a comparison between the performances of different models (including \textsc{alpt-patchy}, \textsc{cola}, \textsc{EZmocks}, \textsc{Halogen}, \textsc{Pinocchio}, and \textsc{pthalos}) at recovering the distribution of BDM halos in the BigMultiDark simulation. In this comparison project one can see that \textsc{alpt-patchy}, \textsc{EZmocks}, and \textsc{cola} yield the most accurate results as compared to the reference simulation in terms of the two and three point statistics (including the quadrupole). Moreover, \textsc{alpt-patchy} and \textsc{EZmocks} were shown to be the less computationally demanding codes using far less dark matter particles than other methods by a factor of 2.37 with respect to \textsc{Halogen} and \textsc{pthalos}, by a factor of 8 wrt \textsc{Pinocchio}, and by a factor of 64 wrt to the original N-body simulation.

\section{Summary and Discussion}
\label{sec:discussion}

This work presents a move forward towards fast and accurate generation of mock halo/galaxy catalogs, extending in particular, the \textsc{patchy} code. We have introduced an efficient MCMC technique to automatically obtain the bias parameters relating the halo/galaxy population to the underlying large scale dark matter field based on a reference catalog. 

This technique is flexible and admits incorporation of different bias models, and number of bias parameters. This permits us to robustly assess the degeneracies and confidence regions of the different bias parameters.

Furthermore we have introduced in the \textsc{patchy} code a particle mesh structure formation model \citep[the \textsc{FastPM} code, see][]{fastpm} in addition to the previous LPT based schemes.

As a demonstration of the performance of this method, we used the halo catalog of the BigMultiDark $N$-body simulation as a reference catalog. Our calibration method makes use of the halo two-point statistics and the counts-in-cells to estimate the bias parameters. 

Based on the dark matter field obtained with \textsc{FastPM}, which includes an improved description towards small scales, and in particular, the enhanced power caused by the one halo term, we have found that previous studies based on generation of the density field with \textsc{ALPT} were overestimating the contribution to the power due to deviation from Poissonity. The density field generated by \textsc{ALPT} has missing power towards the one-halo term regime, which was partially compensated with the deterministic nonlinear bias and partially with higher stochasticity (larger deviation from Poissonity). Though present, this deviation turns out to be less pronounced when the particle mesh gravity solver is used. Also, we have managed to extend the 2.5\% level accuracy of the power spectra from $k\sim 0.3\; \hperm$ to $k\sim 0.6 \;\hperm$, being at the level of percentage accuracy up to $k\sim 0.4\; \hperm$.

We have demonstrated that the novel implementation of the \textsc{patchy} code reaches higher accuracy in terms of the bispectrum towards small scales with respect to LPT based schemes, such as \textsc{alpt}, and even more so with respect to \textsc{EZmocks}, which relies on the Zeldovich approximation. 

The assignment of halo masses must be done in a post-processing step taking into account the underlying dark matter density field. \citet{zhao2015} demonstrated that the mass assignment is more precise when the underlying dark matter field is more accurate (\textsc{alpt} vs Zeldovich). We therefore expect that using \textsc{FastPM} contributes to further reduce the scatter. We leave the investigation of mass assignment for a later work. Analysis of Redshift Space Distortions will also be presented in a future work.


As we have now implemented a PM solver into our approach, we expect that certain high mass range of halos are correctly described and could be found with a friends-of-friends algorithm, the halos which are not properly resolved could be augmented with the method presented here  \citep[see methods to extend the resolution of $N$-body simulations,][]{delatorre,angulo2014,ahn2015}.  

It is important to note that our investigation in this work has been focused on the generation of high mass halo (and subhalo) catalogs. One of the main challenges toward generation of mock galaxy catalogs is sampling of low mass halos. These host fainter galaxies which will dominate the observed galaxy samples in upcoming galaxy survey datasets. 

We leave a thorough investigation of the production of low mass halo catalogs to a future work. This will presumably require more sophisticated bias models including also nonlocal bias terms. The robust, automatic, and efficient methodology presented in this work should be capable of dealing with this.

In summary, the work presented here contributes to set the basis for a method able to generate galaxy mock catalogs needed to meet the precision requirements of the next generation of galaxy surveys.

\section*{Acknowledgments}

We are grateful to David~W.~Hogg, Jeremy~L.~Tinker, Michael~R.~Blanton, Uros~Seljak, Roman~Scoccimarro, and Alex~I.~Malz for discussions related to this work.
MV is particularly thankful to David~W.~Hogg for his continuous support during the completion of this work. FSK thanks Uros~Seljak for hospitality at UC Berkeley and LBNL during January to July 2016. During this time he met MV and was able to collaborate with YF. This work was supported by the NSF grant AST-1517237. GY acknowledges financial support from MINECO/FEDER  (Spain) under research grant AYA2015-63810-P. Most of the computations in this work were carried out in the New York University High Performance Computing Mercer facility. We thank Shenglong~Wang, the administrator of the NYU HPC center, for his consistent support throughout the completion of this study.

The CosmoSim database used in this paper is a service by the Leibniz-Institute for Astrophysics Potsdam (AIP). The MultiDark database was developed in cooperation with the Spanish MultiDark Consolider Project CSD2009-00064. The authors gratefully acknowledge the Gauss Centre for Supercomputing e.V. (www.gauss-centre.eu) and the Partnership for Advanced Supercomputing in Europe (PRACE, www.prace-ri.eu) for funding the MultiDark simulation project by providing computing time on the GCS Supercomputer SuperMUC at Leibniz Supercomputing Centre (LRZ, www.lrz.de).
\bibliographystyle{mn2e}
\bibliography{mybib}
\end{document}